\documentclass[12pt]{spieman}  % 12pt font required by SPIE;
\usepackage{amsmath,amsfonts,amssymb}
\usepackage{graphicx}
\usepackage{setspace}
\usepackage{tocloft}
\usepackage{xcolor}
\usepackage{siunitx}
\usepackage{soul}

\usepackage{lineno}

\title{On-orbit Performance of the {\it Spitzer Space Telescope}: Science Meets Engineering}
\author[a*]{Michael W. Werner}
\author[b]{Patrick J. Lowrance}
\author[c]{Tom Roellig}
\author[a]{Varoujan Gorjian}
\author[a]{Joseph Hunt}
\author[a]{C. Matt Bradford}
\author[b]{Jessica Krick}
\affil[a]{Jet Propulsion Laboratory,
California Institute of Technology, 4800 Oak Grove Drive, Pasadena, CA 91109, USA}
\affil[b]{IPAC-Spitzer, MC 314-6, California Institute of Technology, 1200 E. California Blvd., Pasadena, CA 91125, USA}
\affil[c]{NASA Ames Research Center, Space Sciences Division, Moffett Field, California, 94035, USA}

\cftpagenumbersoff{figure}
\cftpagenumbersoff{table} 
\begin{document} 
\maketitle

\begin{abstract}
The {\it Spitzer Space Telescope} operated for over 16 years in an Earth-trailing solar orbit, returning not only a wealth of scientific data but, as a by-product, spacecraft and instrument engineering data which will be of interest to future mission planners.  These data will be particularly useful because {\it Spitzer} operated in an environment essentially identical to that at the L2 LaGrange point where many future astrophysics missions will operate.  In particular, the radiative cooling demonstrated by {\it Spitzer} has been adopted by other infrared space missions, from JWST to SPHEREx.  This paper aims to facilitate the utility of the {\it Spitzer} engineering data by collecting the more unique and potentially useful portions into a single, readily-accessible publication.   We avoid discussion of less unique systems, such as the telecom, flight software, and electronics systems and do not address the innovations in mission and science operations which the {\it Spitzer} team initiated.   These and other items of potential interest are addressed in references supplied in an appendix to this paper. 
\end{abstract}

% Include a list of up to six keywords after the abstract

% Include email contact information for corresponding author
{\noindent \footnotesize\textbf{*} Michael Werner,  \linkable{michael.w.werner@jpl.nasa.gov} }

\begin{spacing}{1}   % use double spacing for rest of manuscript

\section{Introduction}
\label{sect:intro}  % \label{} allows reference to this section
The {\it Spitzer Space Telescope}, as a long-lived observatory operating outside of the thermal and ionizing radiation environment of Earth orbit, and making use of both modern integrating infrared arrays and radiative cooling, serves as a technical pathfinder for the {\it James Webb Space Telescope} (JWST) and other future astrophysics missions.  Rather than describe the entire {\it Spitzer} system design and performance in detail as found in\cite{gehrz2007, Werne2004}, we concentrate our discussion of the on-orbit performance in areas in which we feel that the {\it Spitzer} experience most uniquely pertains to future missions.  These are detailed in the following sections:

 \renewcommand{\labelitemi}{$\textendash$}
 \renewcommand\labelitemii{$\textendash$}
  \renewcommand\labelitemiii{$\textendash$}
  \begin{itemize}
   \item Section 2. Thermal Considerations
   \begin{itemize}
     \item  2.1 Thermal System Design and Performance – Use of Radiative Cooling
     \item 2.2 The Solar Panel
     \item 2.3 The Solar Panel Shield and the Outer Shell 
     \item 2.4 The Telescope
     \item 2.5 Maximizing the cryogenic lifetime of {\it Spitzer}
     \item 2.6 Integration, test, and verification of the cryo-thermal system
   \end{itemize}
   \item Section 3. Payload Issues
   \begin{itemize}
       \item 3.1 Optical system verification and focus adjustment
       \item 3.2 Instrumental sensitivity
       \item 3.3 Cosmic Ray hit rate
       \item 3.4 Photometric Stability
   \end{itemize}
   \item Section 4. Spacecraft Performance
   \begin{itemize}
       \item 4.1 Electrical Power Generation
       \item 4.2 Pointing System Performance
        \begin{itemize}
            \item 4.2.1  Target Acquisition
            \item 4.2.2 Mapping the sweet-spot
            \item 4.2.3 Improving Pointing Stability
        \end{itemize}
    \item 4.3 Angular Momentum Management
   \end{itemize}
 \end{itemize}

\subsection{Description of the} {\it Spitzer Space Telescope}

\begin{figure*}[h!]
\includegraphics[height=15cm]{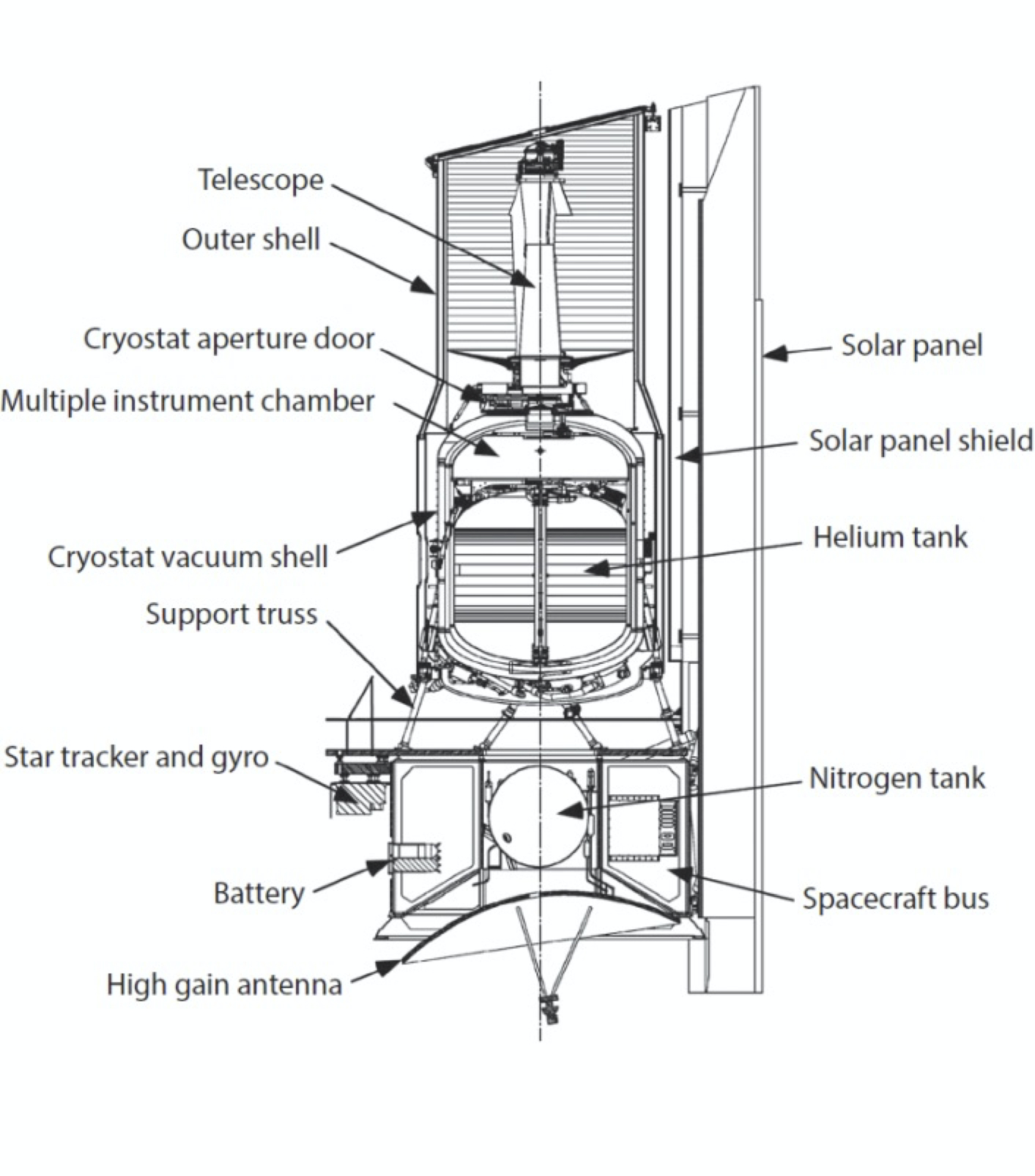}
\caption{Cutaway view of the {\it Spitzer} observatory\cite{Werne2004}. The dust cover atop the telescope tube was jettisoned a few days after launch, and the cryostat aperture door was opened shortly afterwards to admit infrared radiation into the instrument chamber. See papers listed in the appendix for a detailed discussion of the in-orbit checkout for {\it Spitzer}.
\label{fig:Spitzercutaway}}
\end{figure*}

NASA launched {\it Spitzer} into an Earth-trailing solar orbit in August, 2003, as a cryogenic telescope cooled by liquid helium and radiation to space. 
Table \ref{tab:summary} summarizes for the readers the key properties of the {\it Spitzer} mission, and Figure \ref{fig:Spitzercutaway} shows a cutaway view of the observatory as flown.  In considering these properties and the material below, readers should bear in mind that {\it Spitzer} was highly constrained in the 1990s, both in mass/volume by the required Delta launch vehicle and by NASA mandate to keep the cost through launch under \$500M [it eventually grew to about \$750M, including launch vehicle].  These constraints dovetailed nicely with the characteristics of the warm-launch, solar-orbiting mission which emerged as the optimum design solution, providing the most science per dollar.  As exemplified by the discussion below of the Warm Mission, these cost constraints meant that not all conceivable analyses and design trades could be carried out.

\begin{table}[H]
\centering
    \begin{tabular}{lc}
    \hline
    Parameter & Value \\
%    \hline 
    \hline
    Total observatory mass at launch & 861 kg \\
    Dimensions (height$\times$diameter) & 4.5$\times$2.1 m$^2$ \\
    Average operating power & 375 W \\
    Solar array generating capacity at launch & 500 W \\
Nitrogen reaction control gas at launch & 15.59 kg \\
Estimated reaction control gas lifetime & 17 years \\
Mass memory capacity & 2 Gbytes \\
Telescope primary diameter  & 0.85 m \\
Telescope operating temperature & 5.6–13 K \\
(Depending on instrument in use) & \\
Superfluid helium at launch  & 337 L \\
Estimated nominal cryogenic lifetime & 5.6–6.0 years \\
As-commanded pointing accuracy (1$\sigma$  radial) &  $<$ \SI{0.5}{\arcsecond} \\
Pointing stability (1$\sigma$, 600 s) & $\leq$ \SI{0.03}{\arcsecond} \\
Maximum tracking rate & \SI{1.0}{\arcsecond} s$^{-1}$ \\
Time to slew over $\sim$\SI{90}{\degree} & $\sim$8 min \\
Data transmission rate
& 2.2 megabytes s$^{-1}$ \\
(high-gain antenna up to 0.58 AU from the Earth) & \\

Command communication rate & 2 kilobytes s$^{-1}$ \\

    \hline
    \end{tabular}
    \caption{\label{tab:summary} Top-level observatory parameters for the Cryogenic Mission}\cite{gehrz2007}
\end{table}

\subsection{Cryogenic and Warm Missions}

Following depletion of the initial cryogen load in 2009, the {\it Spitzer} telescope and instruments warmed up to the point where only the two shortest wavelength imaging channels remained operational. {\it Spitzer} operated in this  Warm Mission mode very successfully until the mission was  retired in early 2020.  The development team realized from the first that the cryo-thermal architecture, with its heavy reliance on radiative cooling and the use of the telescope outer shell as a thermal boundary, would probably support the Warm Mission.  However, it is important to note that NO resources were expended during the {\it Spitzer} development to optimize or prepare the system for the Warm Mission.   Discussing the Warm Mission would inevitably distract the design/development teams, who were fully occupied with completing the Cryogenic Mission.  In addition, the cost constraints did not make room for significant work on the Warm Mission.  In NASA jargon, the Cryogenic Mission was the Prime Mission, and the Warm Mission was the Extended Mission. The {\it Spitzer} experience clearly shows that the best guarantee of a successful Extended Mission is careful work on the design and development of the Prime Mission.

\subsection{Spitzer's Orbit}  
Unlike an observatory in Earth orbit or at L2, {\it Spitzer} had to address a separate set of challenges arising from the fact that, in the Earth-trailing solar orbit, the spacecraft, launched 25 August 2003, drifted away from the Earth at about 0.1 AU per year, reaching a distance of 1.75 AU at the time of the end of the mission in January, 2020. These challenges, discussed in detail in Ref. \citenum{lowrance2018}, arose from several considerations. The first was the reduced data downlink rate - from 2.2 Mbps early in the mission to 0.55 Mbps at the end - which was a consequence of the increased distance of {\it Spitzer} from Earth. This was compounded by the fact that {\it Spitzer's} fixed solar panel was optimized for solar incidence angles no greater than 30 degrees. During the Cryogenic Mission, the incidence angle was always below 30 degrees, and during the Warm Mission, the incidence angle was always kept below 30 degrees while {\it Spitzer} was observing. However, because {\it Spitzer} downlinked through an antenna fixed to the bottom of the spacecraft, the communication geometry required larger solar incidence angles during the Warm Mission, reaching 55 degrees at the end of the mission. This had implications for battery utilization and recharging and led to solar illumination of structures at the lower end of the spacecraft not originally planned to be in direct sunlight. In addition, elements of the fault protection system had to be over-ridden to allow the use of such large off sun angles without tripping fault protection limits.  This had to be done as part of the setup for each download; following the download the system was restored to its original configuration so that the fault protection would still be operational.   Fortunately, the spacecraft engineering team at Lockheed-Martin, working with the mission operations team at the Jet Propulsion Laboratory and the science planners at the {\it Spitzer} Science Center, found a solution which met all of these constraints, at the cost of some operational flexibility as  only a limited number of data-taking modes were allowed.  Thus {\it Spitzer} was able to operate with efficiency $\sim$90\% up to the end of its mission in January 2020. (Efficiency is defined as (time spent on science, calibration, and slews)/(wall clock time). We will not address these issues in greater detail: We anticipate that the much larger detector arrays anticipated for future missions, as exemplified by JWST, Euclid and Roman, will inevitably point future missions to L2, where none of these problems need arise, and where the thermal and sky-visibility benefits of the solar orbit remain. We did consider L2 instead of the solar orbit for {\it Spitzer} but felt that the mass, cost, complexity and risk (to our cold surfaces) of the station-keeping made it less attractive than the solar orbit  for the heavily constrained {\it Spitzer}.

It is instructive with regard to the previous discussion to note that {\it Spitzer's} orbit had an eccentricity of 0.0113 and a period of 373.1 days, slightly longer than an Earth year. {\it Spitzer} was inserted directly into this orbit by a final rocket burn which ejected it from Earth's orbit. It was this difference to the Earth's orbit which caused {\it Spitzer} to fall further behind the Earth every year. The {\it Spitzer} orbit was chosen both to assure that the observatory was not in danger of falling back to Earth, and,  more importantly, to allow {\it Spitzer} to escape rapidly from the Earth's heat load so that the radiative cooling could proceed.  No station keeping was required in this orbit, and the cold N$_2$ gas system described below was used only for momentum management, not for orbit maintenance.  Finally, we point out that {\it Spitzer} was a robust and reliable spacecraft. Over the 16+ year mission, {\it Spitzer} averaged only slightly more than 1 safing or standby event per year. Fewer than 4 days/yr were lost to these events.

\section{THERMAL CONSIDERATIONS}
\subsection{Thermal System Design and Performance – Use of Radiative Cooling} 

\begin{figure*}[h!]
\includegraphics[height=13cm]{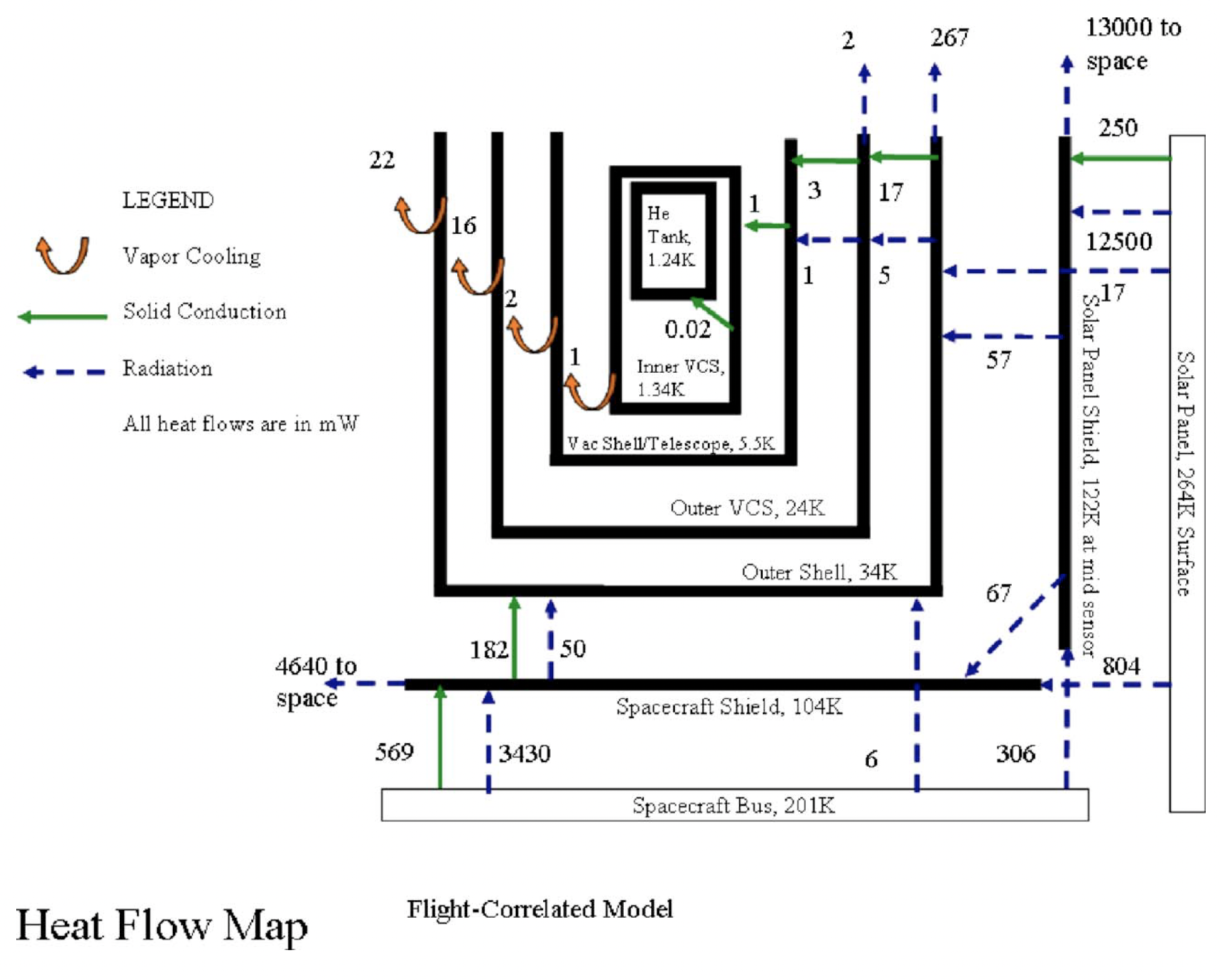}
\caption{Ball Aerospace thermal model\cite{gehrz2007}. Heat input is
solely from insolation on the solar panel. Cooling of the
cryogenic telescope assembly is accomplished by radiation
and vapor cooling. Heat is transferred through the
system along the paths indicated by the arrows by radiation
(dashed blue arrows), conduction (solid green
arrows), and vapor cooling (broad orange arrows). The
equilibrium temperatures for the various observatory
components are given for the case when the cryogenic
telescope is operating at 5.5 K. The model assumes a
focal-plane heat dissipation of 4 mW and an insolation
of 5.3 kW. Courtesy of Ball Aerospace/JPL-Caltech.}
\label{fig:Spitzer-heatflow}
\end{figure*}

The discussion below will refer repeatedly to the heat flow diagram for the cryogenic mission shown in Figure \ref{fig:Spitzer-heatflow}. Paul Finley at Ball Aerospace developed this model, which represents a correlation of the on-orbit performance of the thermal system early in the Cryogenic Mission with the prelaunch thermal model.  Unfortunately, no comparable heat flow diagram exists for the Warm Mission, but some of the elements of Figure \ref{fig:Spitzer-heatflow} bear directly on the thermal performance of the Warm Mission.

Spitzer was launched with its cryostat containing 337 liters (49 kg) of superfluid liquid helium, which cooled the instruments and the telescope.  {\it Spitzer} operated in this Cryogenic Mission mode until mid-2009, utilizing all three science instruments and observing at wavelengths between 3 and 180 $\mu$m.  When the helium supply was exhausted in mid-May of 2009, the system warmed up to a level where only the two shortest wavelength arrays of the Infrared Array Camera (IRAC) instrument, operating at 3.6 and 4.5 $\mu$m, had low enough dark current to be useful for scientific observations.  This phase, referred to below as the Warm Mission, lasted until the observatory was turned off in January, 2020.  Here we discuss the thermal history of key components of {\it Spitzer’s} thermal system over the entire  16+ yr mission.  Figures \ref{fig:Spitzercutaway} and \ref{fig:Spitzer-heatflow} will illuminate this and the following discussions.  Note that, unlike the previous Infrared Astronomical Satellite (IRAS) and Infrared Space Observatory (ISO) missions, which launched with the telescope inside the cryostat, the {\it Spitzer} telescope was launched warm and cooled on orbit, making maximum use of radiative cooling to lose energy to the coldness of space.  

Some idea of the overall efficiency of the {\it Spitzer} cryogenic system is given by the following:  After the loss of helium due to blowdown and on-orbit cooling, the observatory entered its final stages of in-orbit checkout on October 10, 2003, carrying 43.4 kg of liquid helium, down from the 49 Kg present at launch.  This lasted through May 15, 2009, a total of 2030 days, so that the helium was boiled away at a rate of 0.24mg/sec.  The corresponding average heat load to the helium bath was 5.1 mW, consistent with the power dissipated at the focal plane by the instruments and make-up heaters. In comparison, IRAS carried 73 kg of liquid helium and had a lifetime, in low Earth orbit, of 10 months, as opposed to the over 5.5 years achieved by {\it Spitzer} with 42.5 Kg of liquid helium.  IRAS’s helium usage was dominated by parasitic heat conducted and radiated inward from the warm outer shell of the cryostat; {\it Spitzer}, using radiative cooling in the thermally advantageous heliocentric orbit to cool the telescope outer shell, had helium utilization dominated by the much lower power level needed to operate the arrays and the make-up heater used to control the helium bath temperature, and thus the telescope temperature as well (see section 2.5).Although improvements to the {\it Spitzer} thermal design could be contemplated, we emphasize that the heat load to the helium bath was dominated by the unavoidable [though minimal] power demands of the focal plane instruments.  Thus the cryogenic lifetime was in fact determined by the amount of helium remaining after launch and initial cooldown.  The size of the cryostat was limited by the constraints discussed above, so the only design improvements which could significantly increase the cryogenic lifetime would be ones which minimized the amount of helium used in the initial mission phases, and none have been suggested.

\subsection{The Solar Panel}

\begin{figure*}[h!]
\includegraphics[height=10cm]{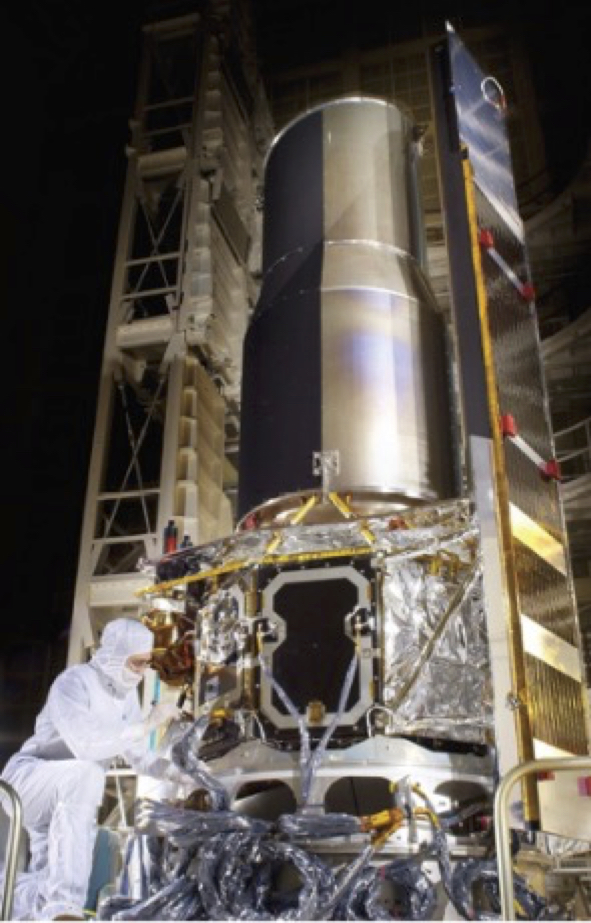}
\caption{The assembled {\it Spitzer} observatory being prepared for thermal vacuum testing at Lockheed Martin in Sunnyvale, CA.  The photograph illustrates the key features of the thermal control system up to and including the outer shell of the telescope. These include the chevron-shaped solar panel, the solar panel shield immediately behind it, and the telescope outer shell, with its anti-solar side painted black to maximize its infrared emissivity.  Also visible between the spacecraft and the outershell is the spacecraft shield which isolated the outer shell from the spacecraft. Credit: NASA.
\label{fig:Spitzerimage}}
\end{figure*}

We discuss the thermal history of key elements of the {\it Spitzer} thermal/cryogenic system, starting at the outside and working inward.  The photograph in Figure \ref{fig:Spitzerimage} shows the elements of the thermal system external to the telescope. {\it Spitzer} was optimized for studies at infrared wavelengths, which required that the telescope, its baffles, and the instruments be cooled to below 10 K.  To achieve this within the cost and mass constraints imposed programmatically, the telescope was designed and operated so as to make maximum use of radiative cooling as illustrated in Figure \ref{fig:Spitzer-heatflow}. During the Cryogenic Mission, the fixed solar panel (cf. Fig \ref{fig:Spitzerimage})  was always oriented to shade the telescope outer shell and the spacecraft from the sun, and the anti-solar side of the outer shell was painted black to enhance its radiative cooling power.   The solar panel assembly was provided by Lockheed-Martin as part of the spacecraft.  Most of the rest  of the flight system including the solar panel shield, the spacecraft shield, the outer shell and the structures within it, as well as the Infrared Spectrograph (IRS) and Multiband Imaging Photometer (MIPS) instruments, was  provided by Ball Aerospace.  Goddard Space Flight Center provided the third instrument, the Infrared Array Camera (IRAC) which was the only instrument in use during the Warm Mission and features prominently in this paper.

To understand the thermal performance of {\it Spitzer}, we start at the solar panel, which absorbs and redistributes all the power – in the form of both electrical and thermal energy – coursing through the 
Spitzer system.  The solar panel was divided into two sections.  The lower portion, consisting of two $330\times72$ cm panels arranged in a chevron arrangement (Figure \ref{fig:Spitzerimage}) was 60\% covered with 11 mil GaAs/Ge solar cells with 18\% efficiency at launch, similar to those which were flown on the Iridium spacecraft. The solar cells provided more than the   $\sim$375W average operating power required by {\it Spitzer} throughout the mission. An aluminum upper section extended the chevron an additional $\sim$160 cm above the active section of the solar panel and shaded the upper section of the telescope outer shell from direct sunlight.  Thus the solar panel served as both sunshade and electric power generator. The sun-facing side of this extension and the regions of the lower section not occupied by solar cells were covered with second-surface aluminized Teflon with an absorption/emissivity ratio of 0.26/0.80, designed to minimize the temperature of the solar panel.

\begin{figure*}[h!]
\includegraphics[width=15cm]{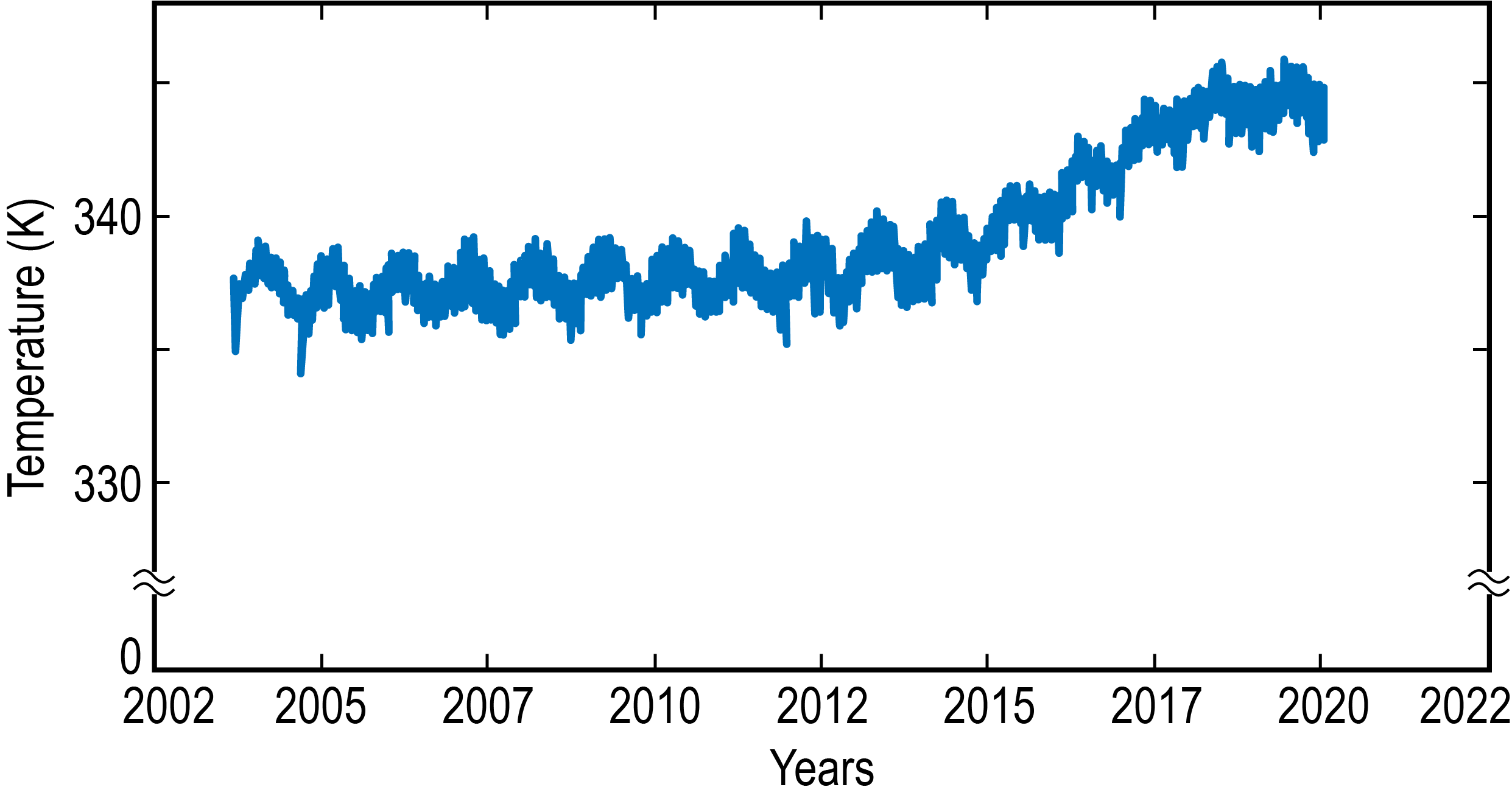}
\caption{The average temperature of the lower, solar cell-containing portion of the solar panel vs. time over the entire {\it Spitzer} mission.  The annual periodicity seen in this and the other temperature data reflects the eccentricity of {\it Spitzer’s} orbit.
\label{fig:solar-panel-temp}}
\end{figure*}

Figure \ref{fig:solar-panel-temp} shows the average temperature of the active lower segment of the solar panel as a function of time throughout the {\it Spitzer} mission.  The temperature shows a gradual increase in time, superposed on a periodic annual variation due to the eccentricity of {\it Spitzer’s} orbit.   Over the 16+ year mission the temperature of the active portion of the solar panel increased from 338 to 344 K, corresponding to an increase in absorbed power by $(344/338)^4-1 = 0.073$, or 7.3\%, because the solar panel cools principally by radiation. This increase is attributed by Ref.~\citenum{finley2005} to “unsurprising degradations in thermal coatings on the solar panel”.  Similar increases in temperature are seen in other long-lived space systems; and attributed, as suggested above, to degradation of the solar panel materials by the effects of unfiltered solar ultraviolet light and the energetic particle environment in space.  In the case of {\it Spitzer}, the increased solar panel temperature is accompanied by a drop in the electrical output (discussed below) which may also be attributed to degradation of the optical properties of the solar panel materials.

\subsection{The Solar Panel Shield and the Outer Shell}

The solar panel shield lies between the solar panel and the outer shell described below (see Figure \ref{fig:Spitzerimage}).    Structurally, the solar panel shield and the outer shell are constructed of lightweighted aluminum honeycomb cores with thin face sheets. The solar panel  shield and the solar panel are stood off from the spacecraft by 
Gr/CE struts and not physically attached to any portion of the telescope outer shell.   The surface finishes are chosen to control heat flow – low emissivity surfaces, such as the facing sides of the outer shell and the solar panel shield, are coated with co-cured aluminized Kapton.  By contrast, as shown in Figure \ref{fig:Spitzerimage} the space facing back half-cylinder of the outer shell is coated with a proprietary Ball black paint \cite{franck2016}, which gives it very high emissivity even at the low temperatures achieved  on orbit. 
As shown in Figure \ref{fig:Spitzer-heatflow}, the solar panel shield coupled to the solar panel primarily through radiation, with only 2\% of the heat  transferred by conduction.  The solar panel shield loses most of this energy by radiation to space, thereby greatly reducing the heat load on the outer shell. Multilayer insulation [MLI] thermal blankets were used to reduce the temperatures of the solar panel shield and the spacecraft shield (Figure \ref{fig:Spitzerimage}), which were adjacent to the telescope outer shell, to 122K and 104K, respectively.  One blanket lay between the solar panel and the solar panel shield, while the other was positioned below the spacecraft shield. 

All of the energy which found its way into the critical telescope/cryostat structure passed through the outer shell.  Thus the telescope outer shell was a critical node in the thermal system, as it set a thermal boundary for the telescope and the cryostat, which sit within it.   The cryostat contained the liquid helium tank and the instrument chamber, which is thermally coupled to the helium tank and not to the cryostat vacuum shell. On orbit, the heat input to the telescope outer shell, a structure $\sim$3m in length and $\sim$1.2 m in diameter, was about 310 mW, which is one part in 17,000 of the 5.3 kW  striking the solar panel.  
The outer shell was  heated approximately equally by radiation and conduction (Figure \ref{fig:Spitzer-heatflow}).   The radiation was from the solar panel shield and from the spacecraft shield, which lay between the spacecraft and the outer shell.  The conduction was from the mechanical and electrical connections between the spacecraft and the outer shell; the microcables passing inwards to the cryostat and the instruments were heat sunk to the outer shell via its supporting truss of low conductivity gamma-alumina struts.  The outer shell lost energy through radiation to space, and, to a much lesser extent, by conduction inward. Because the radiative and conductive heat loads into the outer shell are comparable, the system  was ``well- balanced'' in a thermal sense. During the Cryogenic Mission, the heat conducted inward was carried away by the last stages of vapor cooling provided by the evaporated helium as it left the system through a low-thrust valve.   It is important to note that the heat loads to the outer shell did not vary dramatically between the Cryogenic and Warm Missions beyond the gradual increase implied by the thermal histories shown in this paper.

\begin{figure*}[h!]
\includegraphics[height=10cm]{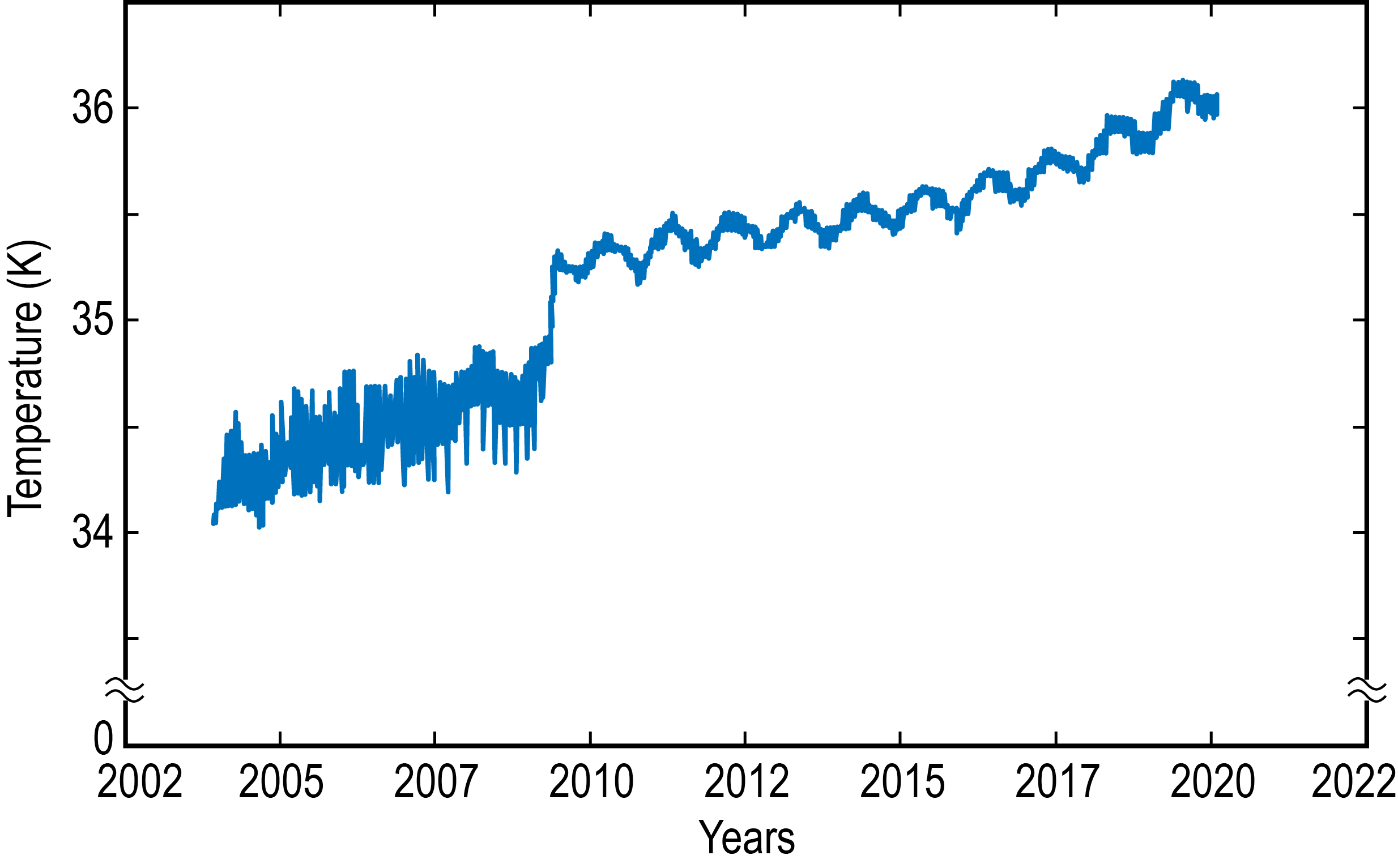}
\caption{Spitzer outer shell temperature vs. time over the entire {\it Spitzer} mission.  The discontinuity in 2009 reflects the exhaustion of the liquid helium and the start of the Warm Mission, when vapor cooling was no longer available. The oscillations in temperature seen before this reflects the variation in vapor cooling resulting from the helium utilization strategy described in the text.
\label{fig:outer-shell--temp}}
\end{figure*}

As the power absorbed by the solar panel increased over the mission,  a corresponding rise occurred in the temperature of the outer shell.  Figure \ref{fig:outer-shell--temp}  shows the outer shell temperature as a function of time over the entire 16+ year {\it Spitzer} mission.  The step upward rise in temperature by about 0.5K seen in 2009 marks the exhaustion of the liquid helium and the start of the Warm {\it Spitzer} mission.   At this point, the absence of vapor cooling allowed the outer shell to warm slightly.  It also allowed more heat to be conducted inwards to the structures within the outer shell. These structures warmed far above the temperatures characteristic of the Cryogenic Mission.  The outer shell temperature  continued to rise slowly after this, as expected from the rising temperature of the solar panel which let more heat into the {\it Spitzer} system.  We also see the annual variation of temperature due to the eccentricity of {\it Spitzer’s} orbit superposed on this long term trend.
The relatively large fluctuations in temperature prior to helium exhaustion in mid-2009 reflected fluctuations in the amount of helium gas evaporating in the cryostat and hence in the cooling power of the escaping helium vapor, and, again, in the temperatures of the structures interior to the outer shell.   These periodic fluctuations result from the helium utilization strategy discussed below.  Starting after helium depletion in mid-2009, however, this cooling path was no longer in effect, but the design and operational features which enabled radiative cooling during the Cryogenic Mission were still in place.  These features kept the outer shell temperature below 36K at the start of the Warm Mission, maintained entirely passively.  

Note that the envelope of the temperature fluctuations prior to mid-2009 show that the temperature rise began immediately after launch\cite{finley2005}.
The increase in outer shell temperature mirrored the increase in solar panel temperature.  The data in Figure \ref{fig:outer-shell--temp} show an increase in temperature from $\sim$35.3 to $\sim$36K during the Warm Mission. In order to isolate the performance of the radiative cooling system external to the outer shell, we extrapolate the data back to the start of the mission and remove the offset due to the loss of vapor cooling. The envelope of the curve suggests perhaps another 0.4 degree increase in temperature would have been recorded. Thus the power radiated by the outer shell in this slightly fictional scenario increased by  $(36/34.9)^4 -1 = 13\%$. Absent a complete thermal model, this can be taken as an upper limit for the degradation in the performance of the radiative cooling alone, because some of the increase must be simply in response to the increased temperature of the solar panel. Note that the outer shell loses energy predominantly by radiation.  In reality, of course, the power radiated by the outer shell increases by closer to 20\% when the effects of the loss of the vapor cooling are taken into account.  Because of the factor of 10 difference in the temperatures of the two systems, this 20\% increase in the power radiated by the cold outer shell is not inconsistent with the 7\% increase in power radiated by the much warmer solar panel.

\subsection{The Telescope}

\begin{figure*}[h!]
\includegraphics[height=10cm]{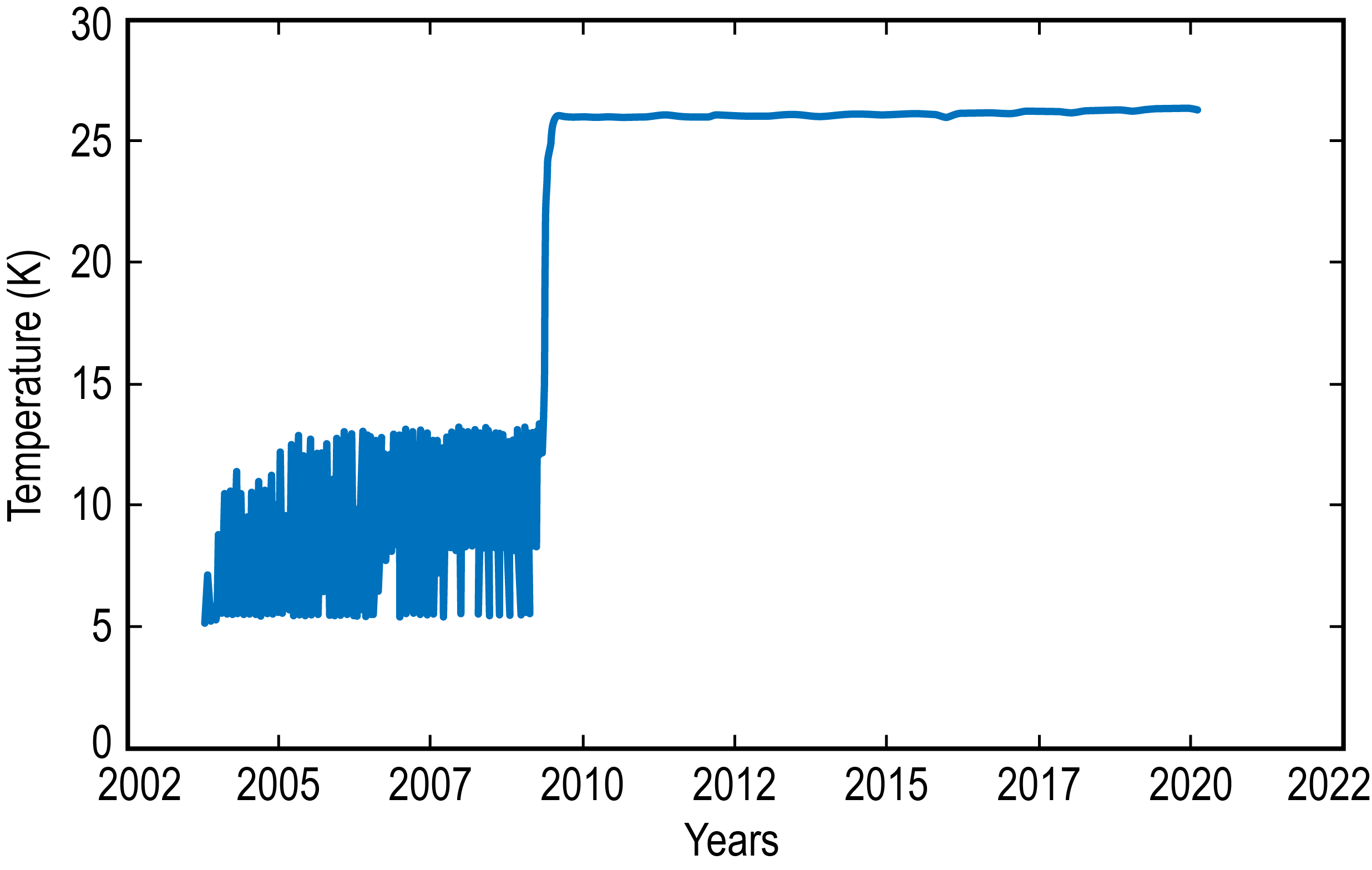}
\caption{Spitzer primary mirror temperature vs. time over the entire {\it Spitzer} mission.  The fluctuations in temperature seen prior to the start of the Warm Mission in mid-2009 reflect the helium utilization strategy described in the text.
\label{fig:mirror-temp}}
\end{figure*}

The telescope optics and metering structure were fabricated of hot isostatically pressed beryllium. Beryllium was chosen because of its favorable strength to weight ratio and its mechanical stability at low temperature.  The optics were polished but not coated. The telescope and its barrel baffle lay within the outer shell and were  surrounded by a vapor cooled shield which intercepted energy conducted and radiated inward from the outer shell.  The telescope was thermally anchored to the exterior vacuum shell of the cryostat which, in turn, was cooled by the helium boil-off.  Two vapor cooled shields provided additional thermal isolation of the helium tank and its associated instrument chamber from the cryostat vacuum shell. The instrument chamber contained the three instruments and the Pointing Calibration and Reference Sensor (PCRS), discussed further below.  As the helium boiloff rate was  varied according to the strategy described below, the cryostat outer shell and the telescope temperatures varied  to meet the needs of the instrument in use.  This  scenario accounted for both the low temperature of the primary mirror and the temperature fluctuations seen during the Cryogenic Mission prior to mid-2009 (Figure \ref{fig:mirror-temp}). Except for the detectors, the instrument hardware was thermally anchored to the helium tank and would rise and fall in temperature very slightly as the bath  temperature changed.  During the Cryogenic Mission, the InSb and Si:As IBC detectors used below 40$\mu$m were thermally stabilized at temperatures above the enclosure temperature, while the Ge:Ga photoconductors in the MIPS instrument, which had to operate below 2K, were thermally strapped directly to the helium bath.

During the Cryogenic Mission, the instrument temperature varied slightly due to the helium utilization scheme described below.  No effects on the instrument optical performance were seen as a result of these changes.  More tellingly, because of the thermal stability of beryllium, it was neither planned nor necessary to refocus the telescope following the transition to the Warm Mission, during which the telescope temperature shifted upward from $\sim$10K to $\sim$25K.\cite{Carey-etal-2010} Over a larger temperature change, it was noted during IOC that the position of the telescope focus stabilized to within 0.01 mm once the telescope temperature fell below $\sim$50K.  

During the Warm Mission, the heat load applied within the cryostat, a combination of power required to stabilize the two IRAC arrays in use and the power dissipated in reading the arrays out, was  constant with time at a level of 1.4 mW. The effects of the  increasing outer shell temperature and its annual variation are just visible in the telescope temperature, which was stable at 26-to-27 K throughout the Warm Mission  (Figure \ref{fig:mirror-temp}).     
The tight mechanical coupling of the telescope, the cryostat, and the barrel baffle suggest that all structures within the vapor cooled shield were at about the same temperature, and these interior structures remained at $\sim$26K by radiating about 20 mW  into space through the open end of the outer shell.   
This suggests that, in addition to the 1.4 mW dissipated within the cryostat by the IRAC arrays, a bit more than an additional 20 mW was conducted or radiated from the outer shell through the shield and into these interior structures.  Although a detailed thermal model for Warm {\it Spitzer} does not exist, this lies in the range of the conductive and radiative loads into the interior of the outer shell seen in the heat flow diagram (Figure \ref{fig:Spitzer-heatflow}). It is important to realize that these loads on the outer shell  should be about the same in the Warm Mission, but they now warmed the telescope  because the cooling effect of the evaporating helium gas was no longer present to counteract them.

\subsection{Maximizng the cryogenic lifetime of {\it Spitzer}}

Spitzer operated only one instrument at a time (typically for a one-week campaign), as appropriate for a system where the cryogen boil-off rate was set by the power dissipated by the instruments within the cryostat.  If two instruments were in use simultaneously, the helium would have been used at a correspondingly higher rate, shortening the mission and reducing the time available for scientists to plan future observations based on initial results.  This single-instrument mode enabled the following strategy for maximizing the cryogenic lifetime of {\it Spitzer}.

The fluctuations in the primary mirror temperature during the Cryogenic Mission (Figure \ref{fig:mirror-temp}) reflect a tailoring of the mirror temperature to the needs of the instrument being used.   Through the use of a heater which increased the rate at which helium was evaporated in the cryostat, the telescope was kept cold enough (5.5K) for natural-background-limited operation when the long wavelength MIPS instrument was in use. The heater was turned off and the telescope temperature was allowed to drift upwards when the shorter wavelength instruments- IRAC and IRS - took over. This tailoring increased the cryogenic lifetime of the system by at least six months over an approach in which the mirror temperature was always kept at 5.5K.  It may have application to future missions which carry expendable cryogens and is described in detail by Ref.~\citenum{lawrence2004} and summarized by Ref.~\citenum{werner2012}.

\subsection{Integration, test, and verification of the cryo-thermal system}

A pre-launch test of the performance of the cryo-thermal system was necessary to assure that {\it Spitzer} would meet its requirement of 2.5 years cryogenic lifetime (the pre-launch goal was five years).  The NASA mantra “Test as you Fly” was, unfortunately, not applicable to tests of {\it Spitzer’s} cryo-thermal system due to the impracticability of replicating the very low thermal background of space, together with solar insolation on one side, in any test chamber large enough to accommodate the entire {\it Spitzer} observatory.  Instead,  we broke this verification into two parts, with the boundary at the outer shell.  The reasoning was simple:  In flight, all heat input into the telescope, instruments, and cryogen system was routed through the outer shell. Thus the telescope temperature and the helium flow rate would depend only on the outer shell temperature and not directly on any external heat sources. The thermal performance of the structures within the outer shell was to be tested in the thermal balance test at Ball Aerospace, carried out with fixed outer shell temperatures.

The thermal performance of the hardware external to the outer shell, which established the outer shell temperature, was evaluated as part of the system level thermal vacuum test, which was carried out at much higher temperatures.  In neither case did the test conditions faithfully replicate the on-orbit conditions, so much of the test effort was devoted to validating the models which were to be used to understand the test results and to extrapolate from the test to the orbital environment.  These tests evaluated not only the design of the thermal system but also the quality of the workmanship involved in its assembly.

Ref.~\citenum{finley2003} describes the low temperature thermal test and verification of the outer shell and its interior components.  The tests were carried out in a chamber at Ball Aerospace with walls cooled by liquid nitrogen, with the telescope, the cryostat, and the instruments assembled for launch within the outer shell (the thermal test was done concurrently with an end-to-end optical test described below).   The outer shell was shielded from a direct view of the chamber walls by extensive thermal shields and blankets. The original intent of the test was to set the outer shell temperature at both nominal and worst-case values expected on orbit and to measure the telescope temperature and the helium flow rate in each case.  The outer shell temperature was to be controlled by circulating cold helium vapor through a cooling loop at the base of the outer shell.  This cooling loop was included in the design for just this purpose, illustrating the importance of outlining the test strategy during the design phase for a complex, difficult to test, system like {\it Spitzer}. 

However, data taken during the initial cool down showed that the heat load on the vacuum shell of the cryostat, which lay interior to the outer shell and was mechanically and thermally coupled to the telescope, was 5 to 10 times the 5-to-10 mW expected on orbit.  This uncontrolled power was small compared to the potential test induced heat input from radiation or solid or gaseous conduction (there were four separate cooling loops bringing in helium from vessels outside the test chamber); for example, a blackbody at 273K radiates about 30 mW/cm$^2$.  
The challenge detailed in Ref.~\citenum{finley2003} was to separate the environmental or test-induced heat loads from those intrinsic to the system; only the latter would be present in flight. To do this, the authors did a careful audit of the possible test-induced heat loads, which are described in detail in Ref.~\citenum{finley2003}, to determine which were credible, and incorporated them into a thermal model of the test configuration which also included the parameters characterizing the thermal performance of the flight system.  They  identified numerous possible sneak heat paths, including, for example, radiation down the pipes used in cooling loops which carried liquid helium from outside the test dewar into the outer shell.   Following correction of these known problems, there were quite a few parameters to be evaluated, and the long thermal time constant of the system made it impractical to test each separately. Instead, a series of 8 separate energy balance cases, each with a different set of environmental and system parameters, were carried out and analyzed in detail. Because the system never reached thermal equilibrium during these tests, much of the analysis focused on the transient behavior of the system, which was predicted by the thermal model. The transient behavior reflected the thermal coupling between various system elements, which was central to the on-orbit performance.   In short, the purpose of the telescope thermal balance test became validation of the thermal model rather than actual demonstration of flight performance.

This painstaking work succeeded in separating the test  artefacts – which were still dominant - and in predicting on-orbit behavior.  It set the stage for the observatory level thermal balance test of the assembled spacecraft, carried out in a large chamber at Lockheed Martin, Sunnyvale, and shown in Figure \ref{fig:Spitzerimage}. As was the case for the test of the outer shell and its interior systems, described above, it was not possible during the observatory thermal balance test (which had other objectives in addition to  supporting an estimate of the cryogenic lifetime)  to replicate the boundary conditions which the observatory would encounter in space. Instead, the analysis focused on three time windows, spread over about three days, during which temperatures of key system elements were measured as the observatory cooled down.  Prominent among the elements investigated were the spacecraft shield and the solar panel shield, because a key objective was to test whether these components, which were expected to dominate the radiative and conductive loads on the outer shell during flight, could adequately isolate the outer shell from the heat of the solar panel and the spacecraft.  Of course, the outer shell temperature was monitored as well.  In a similar spirit to that of the outer shell test described above, the analysis focused on the changes in temperature from one time window to another rather than on the actual temperatures, which were greatly influenced by environmentally-induced heat loads.

Comparison of the actual flight performance with the prelaunch predicts based on the tests described above shows that the latter were remarkably accurate.  The analysis\cite{finley2004} of the prelaunch data, following the thermal balance test, showed that even the worst case (when all thermal parameters were stacked up with their least favorable values), yielded a cryogenic lifetime of 3.1 years, considerably longer than the 2.5 year requirement, while the nominal cryogenic lifetime was predicted to be 5.1 years. This lifetime analysis assumed that the telescope was constantly cooled to its lowest required operating temperature of 5.5K.
 
 It was already apparent that the adaptive cryogenic utilization scheme described above would increase the lifetime by 5 to 10\%, leading to a more realistic lifetime prediction of 5.4 to 5.6 years.  These pre-launch predictions agreed extremely well with the actual on-orbit cryogenic lifetime of just under 5.7 years. Similarly, the predicted temperatures\cite{finley2003} of specific components during the Cryogenic Mission agreed very well with the on-orbit data.  Most importantly, the outer shell temperature was predicted to be 32K, very close to the observed value of 34-to-34.5K (Figure \ref{fig:outer-shell--temp}). 
 
 Finally, when the thermal model was adjusted following launch to fit better the actual temperatures measured during the Cryogenic Mission, it predicted temperatures during the Warm Mission of 24K for the telescope and 36K for the outer shell (P. Finley, private communication).  Again, these predictions were in good agreement with the measured values shown earlier (Figs \ref{fig:outer-shell--temp} and \ref{fig:mirror-temp}).  It is noteworthy that almost 20 years ago the state of the art thermal modelling led to very accurate predictions of the on orbit behavior of a complex system which utilized both radiative and cryogenic cooling.  The performance and predictability of {\it Spitzer’s} radiative cooling approach has been important in reducing the risk of the use of radiative cooling in other NASA missions, such as JWST and SPHEREx.

 In summary, the performance of the {\it Spitzer} cryogenic system on orbit was remarkable. It is a tribute not only to the careful design of the system by our partners at Ball Aerospace, but also to the extreme care with which it was assembled and tested.  For lessons learned from this experience we refer the readers to the publications by the Ball group\cite{finley2003,finley2004,finley2005,hopkins2003} as well as the more general lessons learned analysis of Ref~\citenum{Gehrz-etal-2010}. Two lessons which stick out, however, are the following:  1) For a complex cryo-thermal test such as that undertaken for {\it Spitzer}, the design and fabrication of the test configuration and other GSE need as much care and attention as was given to design and fabrication of the flight system; and, 2) It is important to remember that for systems like {\it Spitzer} for which the on-orbit thermal environment cannot be duplicated in the lab, the thermal balance tests can be used to validate the system thermal model, which can then be used with greater confidence to predict the on-orbit behavior.  In this modelling and validation, it may be found that tests constructed and instrumented to permit analysis of the transient behavior of the system are equally valuable, and considerably easier to implement, than those which rely on reaching a steady state temperature, which may take a prohibitively long time.

\section{Payload Issues}

\subsection{Optical System Verification and Focus Adjustment}

Spitzer used an all-beryllium 85-cm diameter f/12 Cassegrain telescope which illuminated a focal plane $\sim$30 arcmin in diameter.  The focal plane housed pickoff mirrors which fed the modules of all three {\it Spitzer} instruments as well as the PCRS. The telescope was body pointed to place the target of interest onto the pickoff mirror, and hence the entrance aperture, of the instrument/module to be used for a particular observation.  This design required that the instrument modules be confocal; this was achieved by design and test on the ground and verified on-orbit.  
The telescope with the instruments installed underwent optical test and verification at cryogenic temperature in the large cryogenic test chamber at Ball Aerospace.   These tests were simultaneous and interleaved with the thermal verification tests described above.  A standard double-pass test using a cold flat mounted above the telescope, and a short wavelength infrared source placed at the telescope focal plane by Ball for just this purpose, illuminated the IRAC arrays at 3.6 and 4.5 $\mu$m.  The test validated the end-to-end image quality of the system, discussed further below.  It also allowed the {\it Spitzer} team to use the focus mechanism, which was part of the secondary mirror assembly, to set the focus at the estimated position expected for the zero-gravity post-launch configuration. The primary mirror was polished to have the right configuration for its cryogenic performance by measuring the distortions in the figure at cryogenic temperature and polishing the inverse of the measured deviations into the mirror at room temperature.  Following the warm polishing of the mirror and an initial cryogenic test, one additional polishing cycle was carried out following this approach, after which  a second cryogenic test confirmed that the mirror met specifications.   Further details concerning the design, fabrication, test and performance of the optical system are given by Ref.~\citenum{gehrz2007}.

The approach taken on orbit to determine and then set the optimum position for the secondary mirror may be relevant to future space observatories.  The focus mechanism provided motion only along the optical axis and was robustly designed and electrically redundant.   Nevertheless, there was an understandable reluctance to carry out a traditional focus sweep which would require many activations of the focus mechanism at cryogenic temperature.  Instead, a group led by Bill Hoffmann from the University of Arizona developed a technique for determining focal position by looking at the variation of image quality across the combined $\sim$5 by 10 arcmin field of view of the two short wavelength IRAC arrays at 3.6 and 4.5 $\mu$m.  This technique, described by Ref.~\citenum{hoffmann2003} and summarized by Ref.~\citenum{werner2012}, was validated through a double-blind simulation on the ground\cite{gehrzromana2003} and successfully applied on orbit.  A series of on-orbit measurements showed that the position of the telescope focus had stabilized to better than 0.01mm once its temperature fell below $\sim$50K\cite{gehrz2007}, as was expected from the thermo-mechanical properties of beryllium and the other materials in the mechanical assembly.  At this point, the analysis\cite{hoffmann2003} showed that the telescope focus lay 1.8mm above the nominal focal plane established by the instrument entrance apertures.  A small test move of the focus mechanism verified the direction of motion.  Then a larger motion of the secondary mirror brought the focus position within the required range.   The telescope optics alone provided an image FWHM of $\sim$1.45 arcsec\cite{fazio2004} at the center of the 3.6 and 4.5 $\mu$m arrays.  However, when the effects of the instrument optics and array pixelization are considered, the images in the reduced data have FWHM 1.6-to-2” (\linkable{https://irsa.ipac.caltech.edu/data/SPITZER/docs/irac/iracinstrumenthandbook/5/}).  This image quality has supported {\it Spitzer} studies of highly redshifted starlight from galaxies in the distant Universe.

\subsection{Instrumental Sensitivity}

\begin{figure*}[h!]
\includegraphics[height=11cm]{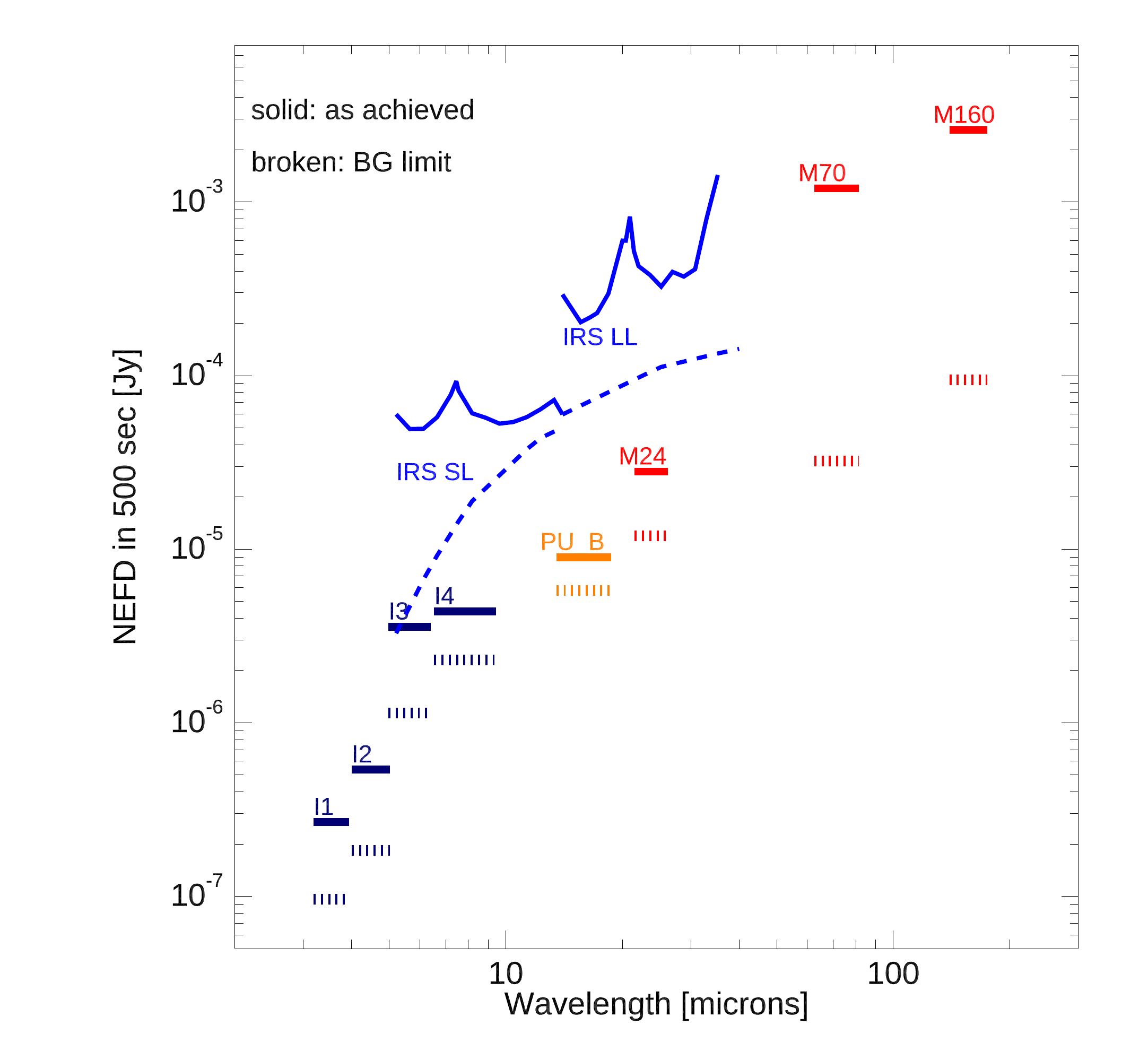}

\caption{Achieved {\it Spitzer} point source sensitivities (solid lines, 1$\sigma$ in 500 sec)\cite{Werne2004} compared with estimates of fundamental limits imposed by the zodiacal background. The short vertical lines under each instrumental bandpass show the zodiacal  background-limited NEFD estimate for the bandpass, assuming noise from the zodiacal light in an annual average sightline viewing the north ecliptic pole. Note that the achieved performance is closest to the background limit in the 8-to-24$\mu$m range were the background is at is brightest.}  The background couples in both polarizations with a square bandpass of the plotted width and total transmission (including detector absorption efficiency) of 0.44, 0.42, 0.14, and 0.30 for IRAC bands 3.6, 4.5, 5.8, and 8.0 $\mu$m,  0.6, 0.18, and 0.15 for MIPS 24, 70 and 160 $\mu$m, and 0.65 for the IRS blue peak-up array (PU B) at 16 $\mu$m. Perfect detectors which add no noise other than generation-recombination noise are assumed throughout. The sensitivity achieved by MIPS is degraded below the zodiacal limit, in part, by the effects of confusion at both 70 and 160 $\mu$m; it is appreciable in both channels in 500 sec.   The IRS short-low and long-low module sensitivities are referred to a $\lambda / \delta \lambda$ = 50 bin; the estimates adopt wavelength-independent efficiencies of 12\% and 8.5\%, respectively; this is meant to include all sources of loss including slit coupling, blaze efficiency in both polarizations, filters and detector quantum efficiency.   For all, it is assumed that the 14\% obscured 85-cm {\it Spitzer} telescope (A$_{\rm geom}$ = 0.489~m$^2$) couples to a point source with 75\% aperture efficiency.   The $\sqrt{2}$ photo-electron recombination penalty is included for all bands except IRAC 1 and 2, where it does not apply.

\label{fig:sensitivity}
\end{figure*}

 With its very cold optics and careful stray light baffling, the {\it Spitzer} observatory achieved photometric sensitivities close to the limits imposed by natural backgrounds in its broad band imaging instruments. Figure \ref{fig:sensitivity} shows a comparison of the measured {\it Spitzer} instrument sensitivity compared to the natural background limit due to the zodiacal emission in the direction of the North Ecliptic Pole. The calculated natural background limit includes the measured {\it Spitzer} instrumental optical throughput as detailed in the figure caption, but assumes otherwise perfect instruments with noiseless detectors and 100\% on-source observing efficiency. The natural background limit also assumes an ideal instrument with a large number of small noiseless pixels that can capture and properly weight the point spread function. In the case of the {\it Spitzer} instruments the pixels are larger than this ideal (to reduce the effects of excess noise that could arise from a large number of real pixels) - which leads to excess sky background, and thus leads to less sensitivity compared to a perfect instrument. In practice, even the zodiacal background limit will actually be degraded by galactic cirrus emission and source confusion, especially at the longer wavelengths. This is readily apparent in the figure where the measured sensitivity deviates most strongly in the MIPS long wavelength bands at 70 and 160$\mu$m\cite{dole2004}. We feel that Figure \ref{fig:sensitivity} and the accompanying discussion show that we can build instruments fully capable of exploiting the low backgrounds of the space environment.

%\clearpage

\subsection{Cosmic Ray Hit Rate}

\begin{figure*}[h!]
\includegraphics[height=10cm]{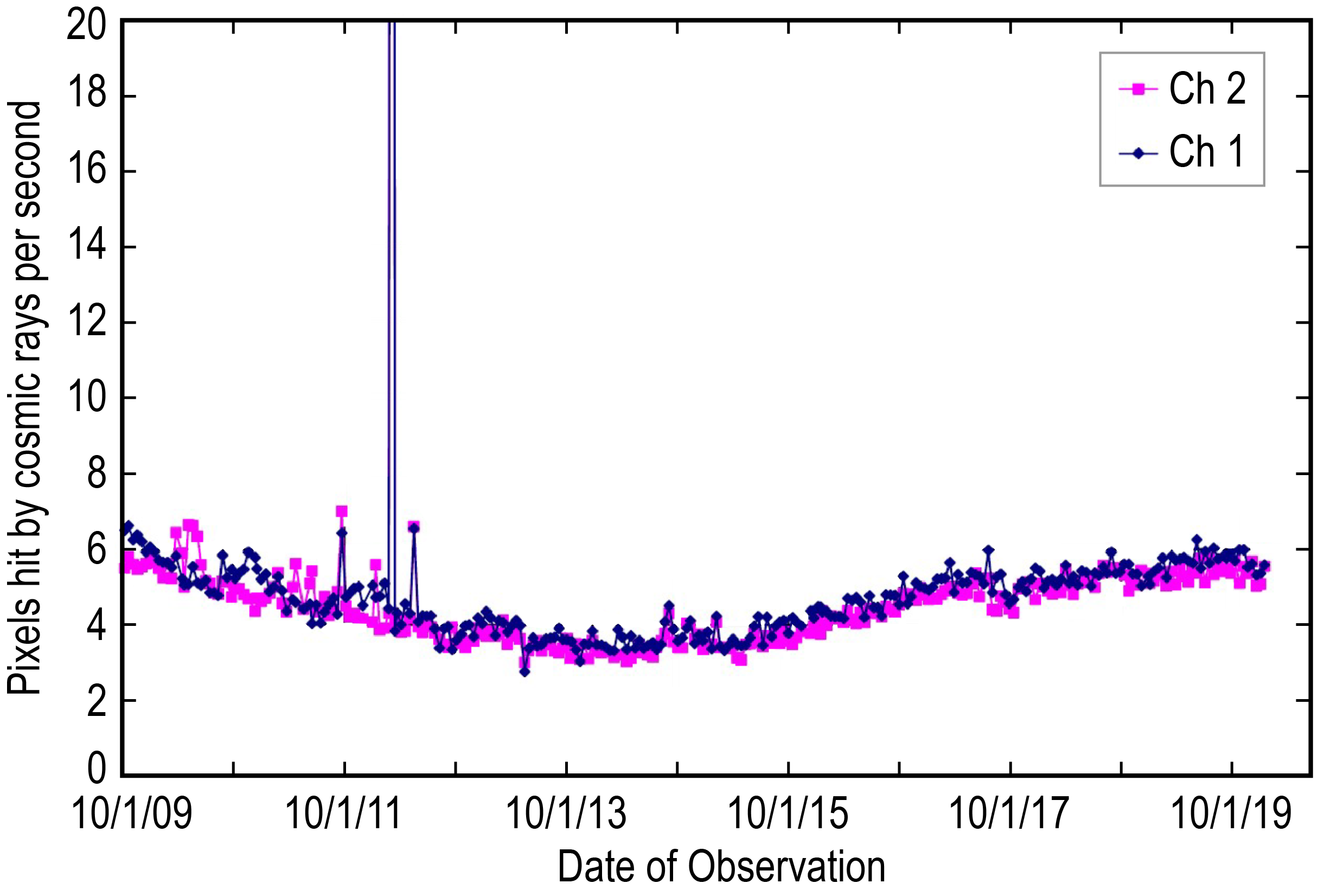}
\caption{Average number of pixels in the IRAC arrays affected by cosmic rays in the calibration frames taken during the {\it Spitzer} Warm Mission\cite{lowrance2018}.  The array formats were 256$\times$256 30$\mu$m-squared pixels,  The spike in 2012 was due to a solar flare.
\label{fig:cosmic-ray-stats}
}
\end{figure*}

Observatories operating outside of the protection of the Earth’s magnetosphere will be more susceptible to the effects of galactic and solar cosmic rays than their better-shielded low Earth orbit cousins.  {\it Spitzer} was designed with radiation-hardened shielding and tested with an assumed radiation environment, and it provided data on radiation effects in the context of a modern observatory using infrared array detectors. We describe here the effects of solar and galactic cosmic rays on the pixels of the 3.6 and 4.5 $\mu$m InSb arrays used by IRAC. For data processing, the number of affected pixels in a known time interval, typically 100 sec, was tracked, using the dark-sky calibration observations taken with IRAC once per week, and was found to correspond to 4–to-6 affected pixels per second (Figure \ref{fig:cosmic-ray-stats}) for a 256$\times$256 pixel array of  $\sim$ 30$\times$30 micron pixels.  The full-well capacity was  $\sim$ 45000 e$^-$  and the array material  InSb.  The threshold for inclusion in Figure \ref{fig:cosmic-ray-stats}  was that the pixels were 10-sigma outliers once all the point sources were removed from a given frame.  Multiple dithers at each pointing position allowed the identification of cosmic ray hits even if they coincided with a point source on the sky.   In most cases, only one pixel was affected at the 10-sigma level by a particular cosmic ray.  The trend observed in the IRAC data followed  the inverse of the solar cycle, as expected, with fewer particle hits during peak solar activity.  The fact that the hit rate contained the imprint of the solar cycle shows that galactic cosmic rays, rather than solar particles, dominated the particle flux except during major solar events.  These images were also used to count the number of bad, noisy, and dead pixels.

Taking 5 pixels/second as a typical hit rate, and bearing in mind that each array consisted of 256$\times$256 30$\mu$m-sized square pixels, and assuming that each incident particle liberated enough charge to register as a cosmic ray hit, the isotropic flux of ionizing particles was about 8 cm$^{-2}$s$^{-1}$.  We feel this is broadly consistent with expectations based on the observed flux of high energy cosmic rays and energetic solar particles. It is noteworthy that the data for each of the two IRAC arrays showed the same hit rate and the same temporal behavior.  This is of course what one would expect if the saturated pixels are attributable to externally-incident ionizing radiation. See Ref.~\citenum{Lowrance-etal-2018b} for more information about the radiation hits and discussion of other issues related to precision Warm Mission photometry with {\it Spitzer}.

The IRAC data processing pipeline monitored all data, even that obtained during  solar flares, to apply corrections for radiation hits before the data were released to the science investigator.  Affected pixels were identified for a single observational period pointed at one part of the sky. This usually consisted  of between 20 and 1000 single frames of the same exposure time. The data processing pipeline created a mask file that kept track of any problem pixels for each frame. As the telescope’s field of view was dithered or scanned across the sky, an astronomical source would be observed in multiple pixels across the array, but a cosmic ray would affect only one pixel (typically) in one frame. When stacked by astronomical coordinates, an astronomical source would appear in multiple frames, so an outlier is a pixel with anomalous flux appearing in only one of the frames. Those outlier pixels were then flagged as radhits, marked in the pixel masks, and excluded when a mosaic, or image map, of the observations was created. This aided in producing artefact-clean images while preserving the processing history.  
Early in the mission, it was found that if {\it Spitzer} were struck by a solar flare and subsequent coronal mass ejection, the number of radiation hits measured by IRAC would rise sharply and quickly fall away. This was used to determine if an observation needed to be retaken. This and other features of {\it Spitzer's} response to high energy particles, including both spacecraft and payload issues, are discussed in reference Ref. \citenum{cheng2014}, which also discusses the effects of the space environment on the solar panel performance. In addition the position of {\it Spitzer} in its orbit allowed data on particle hits as recorded by {\it Spitzer} to be used in partnership with data  from other spacecraft throughout the inner solar system to constrain models  of Solar Energetic Particle events and Coronal Mass Ejections.\cite{palmerio2021,amerstorfer2018}

A figure of merit for the health of the arrays over time is the number of hot and noisy pixels.  These changed with time as some pixels recovered and others became hot or noisy. Pixels were deemed noisy if the standard deviation during the IRAC measurements described above was greater than twice that of the median for surrounding pixels and hot if the pixel retained $>$ 50 counts (DN) after the flux was read out. In Band 1 (the results for Band 2 are similar), the number of hot pixels increased fairly uniformly from 80 to 200 over the course of the Warm Mission, perhaps reflecting the increased accumulated ionizing radiation dose, while the number of noisy pixels bounced around more randomly but generally stayed below the hot pixel count. In practice, we found that it sufficed to update the bad pixel masks about twice per year.  Dead, or non-responsive, pixels increased from about 20 to about 35 during the Warm Mission.  The effects of cosmic ray hits were reversible, and more than 98\% of pixels remained usable at end of the 16+ year mission. A hot pixel table was separately maintained for the star tracker, which used a 512$\times$512 CCD with 20 $\mu$m pixels and a novel “lost in space” acquisition strategy described in detail by Ref.~\citenum{vanbezooijen2003}.  At the end of the mission, there were 46 hot pixels listed in this table.  

Note that at the Observatory level, {\it Spitzer} was protected against cosmic rays by the use of radiation hardened components, including computer, FPGA, and opto-electronic coupler chips.   It is noteworthy that of the 19 events which caused {\it Spitzer} to enter Standby or Safe mode over the 16+ year mission, only one could definitely be attributed to a Single Event Upset (SEU).  This occurred in February, 2009, when an SEU induced a double memory fault in the Combined Electronics unit which served both the IRS and the MIPS Si:As and Si:Sb arrays.

\subsection{Photometric Stability}

The IRAC team monitored the system’s photometric stability at 3.6 µm and 4.5 µm by periodically repeating observations of calibration stars.  The calibration stars included twenty-one primary and secondary stars chosen at the beginning of the Cryogenic Mission\cite{reach2005}.  The calibrator stars were either K giants or A main sequence stars as it was thought that these stellar types could be well modeled and the stars were known to not have astrophysical variations at these wavelengths.  Primary calibrators were located in the continuous viewing zone at the orbit poles so an entire set of primary stars could be observed every two weeks. These were used to determine the flux conversion for each camera. The set of secondary calibrators had positions near the ecliptic plane and therefore only two were visible at any given time.  These were  used to monitor the stability of the photometry over 12-to-24 hr  timescales. 

\begin{figure*}[h!]
\includegraphics[height=10cm]{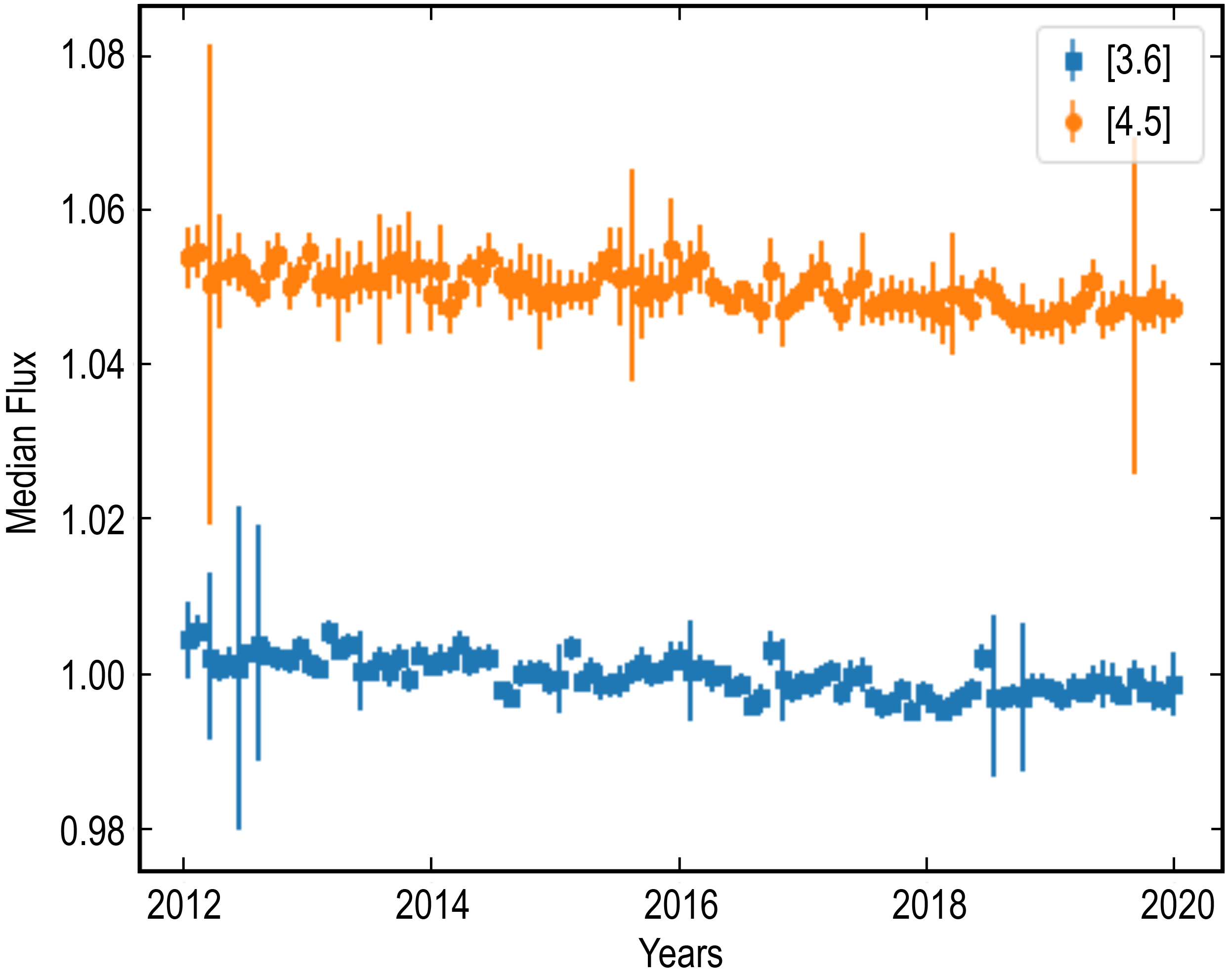}
\caption{Average IRAC responsivity vs. time for the final 8 years of the Warm Mission, based on the measured flux from seven primary calibrator stars. The seven primary calibrators for {\it Spitzer} were KF09T1,KF06T2, KF08T3, KF06T1, NPM1p67, NPM1p60,
1812095\cite{reach2005}.
\label{fig:IRAC-Stability}}
\end{figure*}

We use seven primary calibrators to examine the measured responsivity over the final 8 years of the Warm Mission (Figure \ref{fig:IRAC-Stability}).  The calibration stars were observed with a dither pattern in full array mode. The flux densities were measured with a radius of 3 pixels and a reference sky annulus of 3-7 pixels. They were corrected for both the array location dependence and the pixel phase effect (see below), Dithering to many positions and binning over all stars reduced systematics from each of these effects. Each individual star’s measurements were normalized to the median of all observations of that star over time before being binned together with the other calibration stars at a given epoch to determine the median responsivity at that epoch. Error bars were calculated  for each time-bin by taking the standard deviation of the ensemble  of  photometric points and dividing by the square root of the number of data points in the bin.  The data at channel 1 clustered around a median flux of 1, showing only the small variability discussed below.  The channel 2 data were comparably stable but offset in the plot for clarity. All data were processed with pipeline version S19.2. For further details see Ref.~\citenum{Lowrance-etal-2018b}.

Figure \ref{fig:IRAC-Stability} shows a decline in responsivity of 0.1\% and 0.05\% per year in ch1 and 2 respectively, over the final 8 years of the mission. Similar studies of the Cryogenic Mision\cite{krick2016} show a similar rate of drop in responsivity for the InSb arrays; unfortunately, similar data are not readily available for the other {\it Spitzer} detectors. Other than changes in dark current or noise, which were frequently reversible, there is no evidence of any degradation of the arrays with time.   We feel that the most likely  cause of the drop in responsivity was radiation damage to the optics, which is expected in the space environment and would cause Rayleigh scattering in the transmissive elements.  Other possibilities could be radiation damage to the band-defining filters or to the beam splitters, which were used in reflection  for IRAC bands 1 and 2.  The materials used in these lenses, filters and beamsplitters include MgF$_2$ and ZnS lenses for the 3.6 $\mu$m band and ZnSe and BaF$_2$ for the 4.5 $\mu$m channel.  The radiation from the sky reflected off of a multi-layer dielectric beam splitter with a Germanium substrate, and the band-defining filters were multi-layer coated Ge.  Finally, the arrays were anti-reflection coated with SiO.  For more information, consult the IRAC instrument handbook (\linkable{https://irsa.ipac.caltech.edu/data/SPITZER/docs/irac/}).

The IRAC InSb arrays were thermally stabilized at around 15K during the Cryogenic Mission and 29K for the Warm Mission, and the bias points were reset for optimum performance.\cite{Carey-etal-2010} As a result, the performance of IRAC was degraded by no more than 10\% in going from the Cryogenic to the Warm Mission, and the Warm Mission images are indistinguishable from those obtained during the Cryogenic Mission.  Even at 29K the dark current of the InSb arrays was below the photocurrent due to the zodiacal background, assuring natural-background limited performance (Fig~\ref{fig:sensitivity}). Ref.~\citenum{Carey-etal-2010} discusses how several detector artefacts varied between the Cryogenic and Warm Missions.   For data on the performance of similar InSb arrays operated at higher temperatures in a similar low background environment, see Refs.~\citenum{Baba-etal-2019,Mori-etal-2011}.These papers show that more serious detector artifacts start to appear at temperatures above 40K.

The pixel phase effect alluded to above reflects the fact that, due to intrapixel spatial variations in effective quantum efficiency, the signal received from a star depends on the position of its photocenter on an individual 1.2” pixel. During the Cryogenic Mission, the pixel phase effect could cause signal variations as large as 4\% [peak to peak] at 3.6um but less than 1\% at 4.5um.  For the Warm Mission, the corresponding variations were 9\% and 4.7\% respectively\cite{Ingalls2012}. It can be corrected to first order by using analysis tools available at NASA's Infrared Science Archive (IRSA) which calculate the image centroid and then apply a correction, depending on the centroid position, to estimate the flux which would have been measured if the source centroid was at the position of maximum sensitivity near the center of the pixel.  The correction was based on the (normalized) responsivity measured at different locations within a pixel for hundreds of pixels across the array. For the Warm Mission the correction was a two-dimensional Gaussian centered at the peak response of the pixel (which is slightly offset from the pixel center). The same correction was used for all pixels in a given array.  As will be discussed below, the pixel phase effect was particularly insidious when it was coupled with pointing drifts; even a small motion of the image within a single pixel could produce easily detectable changes in the measured flux.  To be effective, both this correction and that for the array-location variation described below had to be applied to individual data frames prior to any mosaicking or co-adding.

%The average correction is a double gaussian based on the average measured pixel responsivity across the array. 

The array-location dependent correction was required to compensate for the fact that  IRAC was flat-fielded using the zodiacal background, which was not valid for compact sources with star like spectral energy distributions. It also responded to the fact that,as expected, the filter effective wavelength varied across the array due to the change in incidence angle of the radiation from the sky. This effect could  change the inferred flux by 1.3\% on average, with the effect increasing to $\sim$5\% for stars at the edge of the array in IRAC1, and $\sim$8\% in IRAC2. Again, tools made available by the {\it Spitzer} Science Center and available at IRSA correct this effect on a per-pixel basis using a two-dimensional numeric array which also corrects for the small effects of the spatial distortions across the detector array.

\section{Spacecraft Performance}
\subsection{Electrical Power Generation}

\begin{figure*}[h!]
\includegraphics[height=10cm]{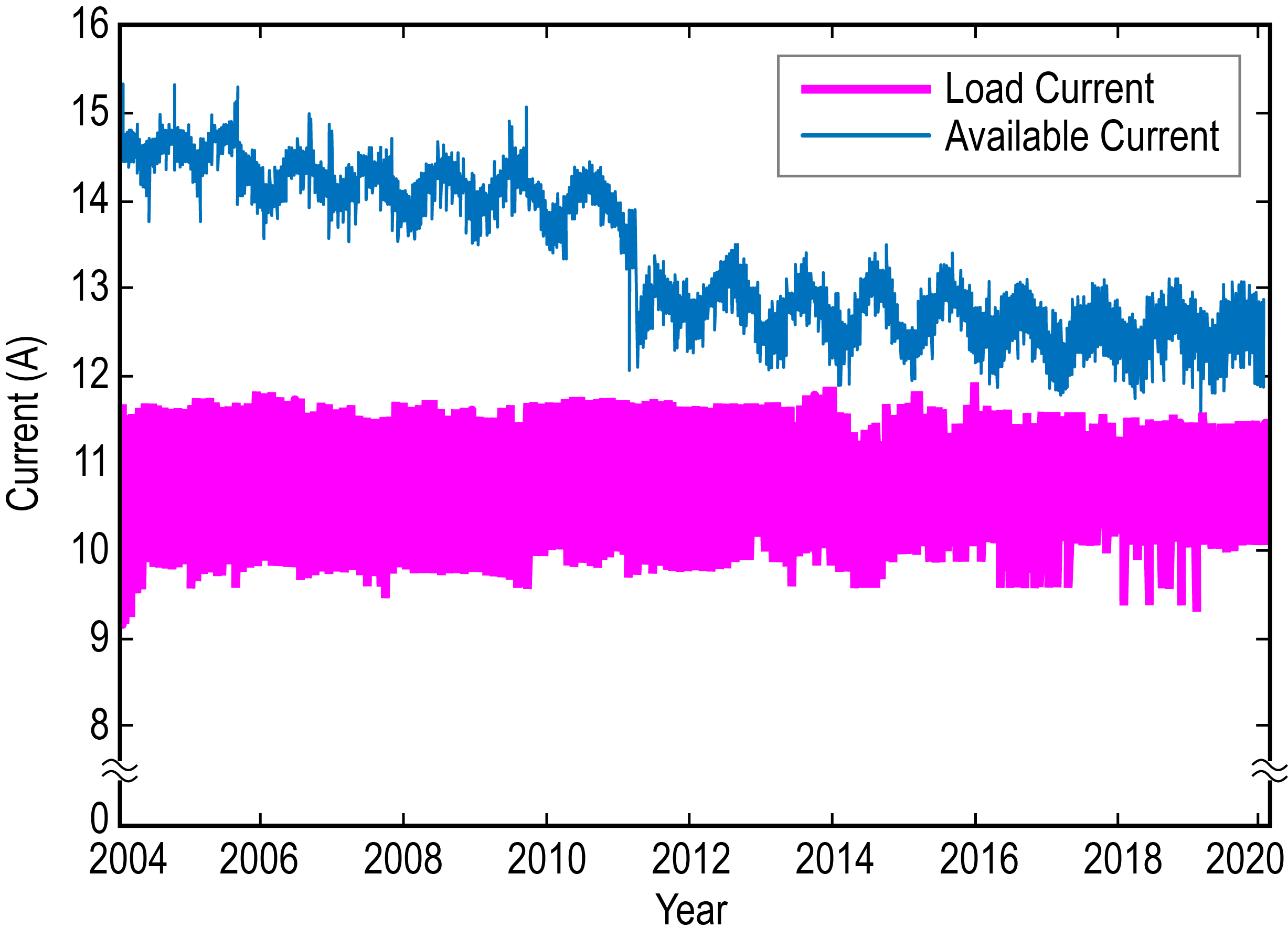}
\caption{Average solar panel-generated current over time for the entire {\it Spitzer} mission.
\label{fig:Avg-Solar-Power}}
\end{figure*}

A degradation in material properties similar to that which led to the increased solar panel temperature shown in Figure \ref{fig:solar-panel-temp} above led to a decrease in the power provided by the solar panels  over time.  Figure \ref{fig:Avg-Solar-Power} shows the average current produced by {\it Spitzer’s} solar panels over  the mission. A gradual decrease in output current is seen with the familiar annual modulation due to the orbital eccentricity.  Also shown is a sharp drop in output in November, 2010, which is inferred to be a micro meteoroid impact which disabled a portion of one of the strings of solar cells in one of the panels.  Even with that loss of capability, however, the average power generated by the solar panels was adequate to meet the needs of the observatory with margin.  

\subsection{Pointing System Performance}

{\it Spitzer} incorporated a sophisticated realization of a standard pointing system with gyros and a star tracker to establish and maintain the orientation of the spacecraft, and a sensor within the focal plane which monitored the coalignment of the telescope line of sight with that of the spacecraft.\cite{Bayard2004} Slews for target acquisition or within an observing sequence were carried out by a set of four reaction wheels.  Thermal considerations required that the telescope was only weakly connected, mechanically, to the spacecraft,  Pointing offsets and drifts between the telescope and star tracker lines of sight were monitored and corrected with the help of  the PCRS, a pair of 4$\times$4 arrays of Si p-i-n diodes which were installed in the cold focal plane and provided visible light images of stars also seen by the spacecraft star trackers. {\it Spitzer's} star-tracker-to-telescope alignment calibration was performed by taking the centroid of a single star in the PCRS sensor once every 8 hours. Each single-star calibration update was processed using a Kalman filter denoted as the S2P (Star-tracker-to-PCRS) calibration filter.\cite{Bayard2007} S2P alignment was accurate to approximately 0.44'' just prior to S2P update, and to 0.35'' just after S2P update (both 1-$\sigma$, radial). This 0.35 to 0.44'' accuracy was maintained throughout the mission. Without the S2P calibration, this critical alignment could drift by as much as 1 or 2 arcsec in 10 days.  The placement of targets on the desired pixel or spectrograph slit was facilitated as well by a precise survey of the {\it Spitzer} focal plane carried out during IOC.\cite{Bayard-etal-2009}

The offsets and drifts were kept minimal by the fact that the telescope and the spacecraft were shaded from sunlight - and thus kept in a stable thermal configuration - by the solar panel and its extension,  In addition, a heat pipe was used to isothermalize the spacecraft top deck, just below the attachment points of the telescope struts, in order to minimize the effect of changes in the thermal state of the spacecraft on the telescope line of sight.  The performance and reproducibility of  the system was further improved by the fact that the two instruments for which system target acquistion was most critical, the IRAC and the IRS, had no moving parts.

During the Cryogenic Mission, the most critical absolute pointing requirements were established by the Infrared Spectrograph (IRS), which required that targets be placed on the spectrograph slits with sub-arcsecond precision.  This was achieved by the use of ``peakup arrays'' within the IRS which provided infrared images of the target so that the telescope could be offset to bring the target from the peakup array on to the spectrograph slit. As many of the science targets were  detectable only in the infrared, the visible light star trackers and PCRS were not useful for this purpose. 
This system performed very well for most purposes, achieving pointing and offset accuracies better than 0.5'' (1-$\sigma$ rms radial), with jitter less than 0.03'' (1-$\sigma$ rms radial) over 600 sec.\cite{Bayard2004} Here we focus on adaptations to this basic system which were driven by the use of {\it Spitzer} to observe exoplanets, or planets around stars other than the Sun, which became a major science theme for {\it Spitzer} within a few years of launch. Exoplanet studies required non-standard approaches to both the understanding and the use of the pointing and control system.

The discovery of the first transiting exoplanets in 2002 led to the realization that {\it Spitzer} was capable of measuring not only the depth of a transit, which occurs when the planet moves in front of the star and blocks its light temporarily, but also the depth of a secondary eclipse, which occurs when the planet moves behind the star so that its infrared light is no longer seen.  {\it Spitzer} observations of secondary eclipses of two exoplanets, published in 2004\cite{Deming-etal-2005,Charbonneau-etal-2005}, were the first detections of light from exoplanets.  In subsequent years, exoplanet studies became an important and growing portion of the {\it Spitzer} science program.  Exoplanet studies required measurement precision better than 100 parts per million over periods of hours, far in excess of anything which had been achieved at {\it Spitzer's} wavelengths from previous ground- or space-based telescopes, including {\it Spitzer}.  {\it Spitzer} users and project personnel worked together over a period of several years to understand how to use both the instruments and the spacecraft, including the pointing system, in previously unanticipated ways to optimize the collection of data on exoplanets, and how to analyze the data to achieve the required levels of precision.

The depth of an exoplanet transit was less than 1\% and often less than 0.1\% of the signal, and so it was crucial to correct for intrapixel variations to obtain any science. Some of the first exoplanet observations employed a dithering strategy (intermittently shifting the position of a target on an array); however, even after the dithering strategy was abandoned in favor of keeping the target fixed on a well-characterized pixel, the main systematic noise limiting the precision of photometry was found to be small changes in pointing position coupled with the intra-pixel gain variations.  Attempts to understand and mitigate these and other known systematics have been accomplished through the following operational changes:  Increasing target acquisition precision and repeatability, independent calibration of the intrapixel gain map, and increasing the overall stability of pointing.  The application of these techniques was validated by several community-wide data challenges\cite{ingalls2016}.

\subsubsection{Target Acquisition}

The intrapixel gain variations could exceed 4\% in Band 2, depending on just where in the 1.2$\times$1.2 arcsec pixel the target centroid fell; they were larger in Band 1.  To mitigate this, a single pixel (the ``sweet-spot'' pixel) at the center of each of the IRAC subarray sections (a 32$\times$32 pixel unit within the larger 256$\times$256 array which was used to achieve higher sampling frequency and make observable brighter targets than could be measured using the entire array) was selected. The PCRS was used to enhance the repeatability of target acquisition by placing the target star on the sweet-spot pixel. The PCRS operated in the visual part of the spectrum (505-595 nm) and its main function was to calibrate and remove the optical offset between the star trackers and the telescope.  In effect it was ``repurposed'' to provide a means of placing optically bright exoplanet host stars on the sweet-spot described below. 

\begin{figure*}[h!]
\includegraphics[height=10cm]{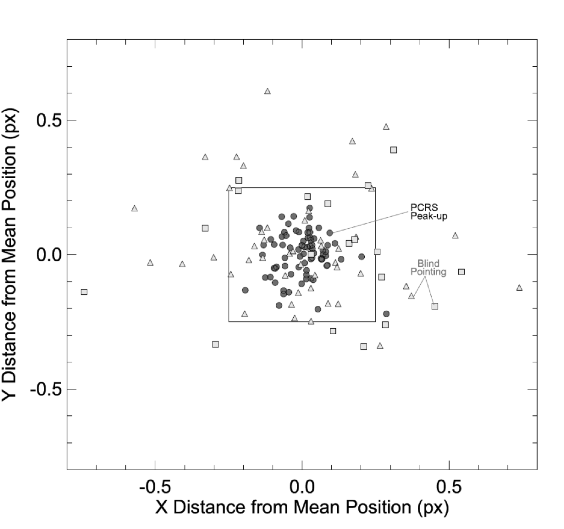}
\caption{Comparison of pointing repeatability using peak-up vs. blind pointing.  Each point represents the mean centroid of the first 64 frames of the following:  78 observations (dark circles) beginning with PCRS Peak-Up followed by an offset to the 4.5 $\mu$m sweet-spot using gyros; 20 observations (light squares)  using PCRS Peak-Up and the star tracker for the offset; and 48 observations (light triangles) initiated  without PCRS peakup. Each set of symbols has had the 
mean position of the set subtracted.  An 0.5$\times$0.5 pixel square is overlaid on the plot, indicating the size of the sweet-spot region\cite{Ingalls2012}.
\label{fig:pcrs}}
\end{figure*}

The PCRS could measure the centroids of stars in the 7.0 mag $<$ V $<$ 11.8 mag range to within 0.14'' (1-$\sigma$ radial) and feed these data, in the form of a pointing correction, to the spacecraft pointing control system. The PCRS Peak-Up repeatedly placed a target within an 0.5$\times$0.5 pixel sweet-spot on the designated 1.2'' pixel (see Figure \ref{fig:pcrs}). The highest precision was obtained when the peak-up star was also the planet-bearing star which was the target of the observation. For fainter host stars, the peak-up was done on a brighter nearby star, and the system performance degraded slightly. When the Peak-Up target was the science target, more than 90\% of visits were placed somewhere inside the sweet-spot; 68\% of visits were within 0.1 pixels of the center of the sweet-spot; and repeat visits to the same target resulted in a spread of initial positions less than 0.06 pixels 68\% of the time. This gave two advantages: (1) it minimized the offset in intrapixel gain caused by initial target acquisition; and (2) it allowed us to map out the responsivity of a small portion of the designated pixel with high fidelity to facilitate data analysis.

For long exoplanet observations, particularly during the latter phases of the Warm Mission, {\it Spitzer} was susceptible to long term pointing drifts following downlinks, as the system relaxed thermally after the heating of the spacecraft during the off-sun excursions.  Coupled to the intrapixel gain variations, these could become problematic when lengthy exoplanet observations, which could extend over 24 hours or more, were executed.  Investigation of the pointing of {\it Spitzer} on a range of time scales showed that on top of the thermal relaxation there was a roughly constant, long-term drift of 0.34 arcsec per day.\cite{Grillmair-etal-2014} This drift was due to an inconsistency in the way that velocity aberration corrections were handled by the spacecraft's Command and Data Handling computer (C\&DH) and by the star trackers. Whereas the C\&DH assumed that velocity aberration would be constant over the course of each observation, the star trackers updated the velocity aberration corrections continuously. The result was a constant error signal sent by the star trackers, which the spacecraft constantly attempted to correct, pulling the line of sight away from the initial pointing position. The strategy adopted to deal with the pointing drifts was twofold: The first was a 30 min ``cooling off period'' following a return to the observing orientation before any pointing-critical observations were carried out.  During this time period the co-alignment of the telescope and the spacecraft relaxed to the usual subarcsecond level  This was augmented by repeated executions of the peak-up procedure, carried out about every 12 hours, which led to a repointing and dealt with all long term drifts.

\subsubsection{Mapping the sweet-spot}

The availability of the PCRS peak-up capability facilitated detailed mapping of the gain variation around the sweet-spot.  In a typical 12 hr time series measurement, the coupling of pointing fluctuations and intra-pixel gain variations could  cause measured fluxes to vary by more than several percent.  Usually these variations were removed from the science data (“self- calibration”) by decorrelating the systematics from the signal using quantities, such as x and y centroid, that changed in tandem with pointing fluctuations and thus were independent of astrophysical variations.   However, the  aliasing of astrophysical and instrumental variations was  always a possibility and could not  be ignored without either an accurate model of the variations of the astronomical system under study (which would limit observers to previously studied systems), or of the instrumental gain. Thus, to permit an independent estimate of the position-dependent instrumental gain, the IRAC instrument science team accumulated a set of approximately 1 million repeated photometric observations of non-variable calibration stars in channels 1 and 2 over the sweet-spot region of each array. The data set required about 150 hours of observatory time.  The responsivity maps\cite{Ingalls2012} show that, even over the 0.5$\times$0.5 pixel central region of the sweet-spot, the gain in IRAC band 1 varies 3-to-5\%. The use of the sweet-spot gain map in analyzing exoplanet time series data was one of the data reduction schemes compared by Ref.~\citenum{ingalls2016}.

\subsubsection{Improving Pointing Stability}

\begin{figure*}[h!]
\includegraphics[height=10cm]{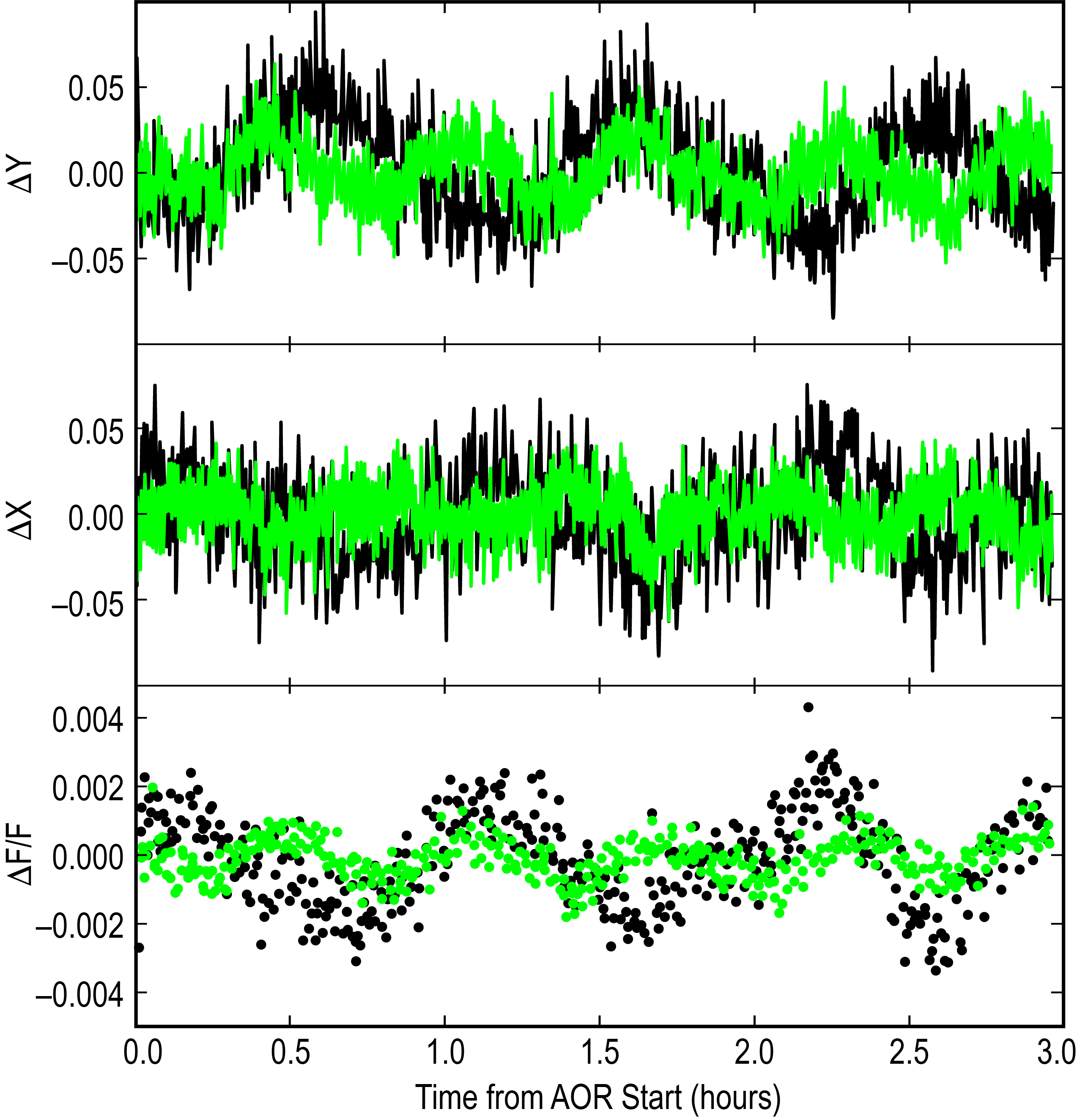}
\caption{Effects of tightening battery heater dead band on pointing and photometric stability.  The upper two panels show the variation of the (x,y) position of the centroided stellar image on the array, while the lower panel shows the variation in measured signal, all as a function of time.  Black points are data taken before the dead band was tightened up; green points were taken after the adjustment.
\label{fig:pointing-drfit}}
\end{figure*}

Soon after scientists started systematic exoplanet studies with {\it Spitzer}, it became apparent that there was a persistent modulation of the signal seen by IRAC with a period of about one hour and a peak-to-peak amplitude of 0.4\% (Figure \ref{fig:pointing-drfit}). This could readily be seen in the time series photometry of measurements of bright stars.  This one hour period was inconveniently close to the length of many of the exoplanet transit and eclipses being studied by {\it Spitzer} users, and the amplitude of the modulation was large compared to the precision required for many {\it Spitzer} exoplanet observations.  A search of spacecraft engineering records revealed that a battery heater was cycling on and off with just this period, which was established in large part by the dead band over which the temperature was allowed to drift before the heater state changed autonomously. It was suspected that the change in heater power changed the thermal state of the spacecraft, introducing a slight offset between the star tracker and telescope lines of sight.  Because the pointing position was established by star tracker observations, this, in turn, led to a change in the observed signal due to uncorrected intra-pixel gain variations.  The 0.4\% signal modulation initially reported, corresponding to an 0.1 pixel  peak-to-peak pointing oscillation, was large compared to the 0.01\% (one hundred parts per million) precision required for many of {\it Spitzer’s} exoplanet studies.   In response, the {\it Spitzer} engineering team tightened the heater deadband so that it cycled on a period of approximately 40 minutes leading to a pointing fluctuation of $\sim$0.05 pixel  peak to peak and a signal modulation of 0.2\% (Figure \ref{fig:pointing-drfit}). Separating the instrumental and exoplanetary timescales and reducing the amplitude of the pointing fluctuations led to increased photometric precision by facilitating more precise subtraction of this signal variability as part of the data analysis.  The close partnership between the {\it Spitzer} users and the {\it Spitzer} Science Center facilitated the reporting and tracking down of this troublesome effect, as it was much more apparent in the science data than in the engineering data from the pointing system.   

\subsection{Angular Momentum Management}
Prior to launch, it was recognized that the pressurized nitrogen gas that was used with {\it Spitzer's} cold gas thrusters to unload angular momentum accumulating in the reaction wheels was a potential life-limiting consumable. The center of pressure vector from the solar radiation impinging on the observatory did not pass through  the center of mass and so solar radiation pressure produced a net torque around the center of mass. As the Helium cryostat was located close to the Observatory center of mass, boil-off of the He liquid had minimal effect on this offset over the course of the mission. (The difference in the location in the observatory center of mass between a full and an empty cryostat was less than 5 cm.)  The torque due to the radiation pressure had to be counteracted with the reaction wheels. If the observatory spent too long in one orientation this momentum built up in the wheels and would have to be unloaded using the cold gas thrusters, expending some of the onboard nitrogen gas each time.  (By contrast, in low Earth orbit this can be accomplished by magnetically torquing against the Earth's magnetic field). Note that the thrusters were not intended to provide a delta-v to the observatory, only to provide a rotational torque to unload the reaction wheels.

 If observations were planned so that equal time was spent looking down as looking up with respect to the orbital plane, the momentum build-up in the wheels due to the solar radiation pressure in one orientation could be canceled.  As it turned out, this was not a major driver in planning observing sequences - optimizing the time spent on observations of the science targets was more important.    Of greater consequence was the fact that, as with time the momentum buildup was understood, it became possible to schedule unloading of the stored angular momentum at a benign time to avoid disturbing the spacecraft with an autonomous desaturation at a critical time, such as during a long integration on a target. At the end of the mission in January, 2020, less than 60\% of the 15.6 kg of N$_2$ carried into orbit had been used. Thus there was some conservatism in the reaction control system design, which was appropriate since the rate of angular momentum build-up would depend on the exact observing scenario.  There might have been some science observations proposed that would have led to a greater imbalance in the observing  time spent looking ``up'' vs. ``down.'' As it was, there were 1066 desats (with apologies to William the Conqueror) during the 2030 day {\it Spitzer} mission.  Thus on average it took about two days for the angular momentum to build up in the reaction wheels to the point where it tripped preset limits for desaturation or a desaturation was commanded in advance of a long and critical observation.

\section{Lessons Learned and Conclusions}

In 2010 co-authors Werner and Roellig collaborated with Robert Gehrz on an SPIE publication discussing lessons learned from {\it Spitzer}\cite{Gehrz-etal-2010}. These included scientific, technical, and management issues but were necessarily weighted towards the Cryogenic Mission.  We describe below several additional lessons which have risen to the top as we look at {\it Spitzer} through the lens of the Warm Mission:
\begin{enumerate}

\item {\bf Radiative Cooling.} 
{\it Spitzer's} innovative use of radiative cooling has been particularly significant. From the point of view of evaluating the durability of the radiative cooling alone, we can say that after 16+ years at 1 AU from the sun, a well-designed and well-constructed radiatively cooled system maintained the telescope temperature at 26K to within a fraction of a degree. Stepping outward, the temperature of the outer shell increased by no more than 1.5K over the 16+ years, and much or all of that can be attributed to the increased power absorbed by the solar array, which trickled down to the outer shell. So there is no evidence over this time period of degradation in the low temperature thermal and optical properties of the components of the system responsible for the radiative cooling.  It is also noteworthy that the cryogenic lifetime of {\it Spitzer} was predicted to better than 5\% from prelaunch models and the results of thermal balance tests on the ground.  In Section 2.6  above we list several other ``lessons learned'' relevant to the cryo-thermal system.  In closing, we emphasize once again that the success of the {\it Spitzer} cryo-thermal system is a tribute not only to the quality of the design but also to the skill and care with which the system was assembled and tested by the Ball team.

\item{\bf Extended Mission.} Several of our thoughtful referee's reports asked what had been done during the Cryogenic Mission, which was the ``Prime Mission'' for {\it Spitzer}, to prepare or optimize for the Warm Mission, which in NASA jargon is an ``Extended Mission.''  Our answer was simple:  Nothing.  We feel that it would be a bad idea to get distracted during the development of a system for its Prime Mission to focus any resources or energy on considerations of a possible Extended Mission.  The best guarantee of a successful Extended Mission will be a successful Prime Mission.

\item {\bf Teaming is Important.} There are two sides to this.  {\it Spitzer} benefited immeasurably from good relationships with our major industrial contractors, Ball Aerospace and Lockheed Martin.  {\it Spitzer} management took every possible step to integrate these contractors into all appropriate activities while empowering them to use their own processes and methodologies.  {\it Spitzer} scientists visited the contractors periodically over the years to discuss {\it Spitzer} science and build additional bridges to the contractors.  As a result, the contractors were fully engaged during the mission, and, in particular, we had excellent support from the Observatory Engineering Team at Lockheed Martin during the Warm Mission when it became necessary to use the spacecraft in previously unanticipated ways.  At the other end of the chain, we found that close relationships with our user community were important in maximizing the scientific output of the observatory.  Perhaps the best example of this comes from the exoplanet investigations, where issues with the pointing and control system at a scale finer than the PCS sensors could detect were identified by {\it Spitzer's} users in the scientific data, and brought to the attention of the Observatory staff.  Future missions should remain alert to the possibility that important engineering information may be gleaned more easily from the scientific data than from the spacecraft telemetry.

\item {\bf Be alert to the possibility of repurposing existing flight hardware.}  The best example of this is the use of the PCRS for precise target acquisition and positioning during exoplanet phase curve measurements.  This was not envisioned when the PCRS was designed to track pointing offsets between the telescope and the spacecraft.  A  slightly quirky example of this was the use of data from {\it Spitzer} to set limits on the characteristics of Coronal Mass Ejections which was made possible by its unique position in the Solar System.

\end{enumerate}

We hope that this distillation of the {\it Spitzer} experience demonstrates the remarkable engineering performance of the {\it Spitzer} Space Telescope and that it  will be of use to colleagues working on future astrophysical observatories in space. We have limited the discussion to areas where the {\it Spitzer} experience is novel and/or likely to be of general interest. In doing so, we have passed over a number of topics where {\it Spitzer's} performance has been noteworthy and which may in themselves contain lessons for future missions. These include operational innovations which are not touched upon in this paper, Many of these other topics are described in a series of papers listed in the Appendix.

\appendix

\section{Bibliography of {\it Spitzer} resources with an emphasis on mission science operations}

\subsection{General Archives}

\subsubsection{Spitzer Documentation and Tools}
Spitzer technical information, including instrument and telescope handbooks, etc.:

\linkable{https://irsa.ipac.caltech.edu/data/SPITZER/docs/
}
\subsubsection{{\it Spitzer’s} scientific data}

Spitzer Heritage Archive:
\linkable{https://irsa.ipac.caltech.edu/Missions/spitzer.html}

\subsection{An overview of {\it Spitzer} science}

\subsubsection{}
“More Things in the Heavens”, by Michael Werner and Peter Eisenhardt, published by Princeton University Press (2019).

\subsubsection{}
“A Spitzer retrospective,” Nat. Astron., 4 293 (2020). \linkable{https://doi.org/10.1038/s41550-020-1089-0}

\subsection{Papers of General Interest}

\subsubsection{}
Gehrz, Robert D, TL Roellig, and MW Werner, 2010, {\it Writing a success story: lessons learned from the Spitzer Space Telescope, An Optical Believe It or Not: Key Lessons Learned II}, International Society for Optics and Photonics, 20 August 2010, 7796, 779602, doi:10.1117/12.864363.

\subsubsection{}
Lowrance, Patrick, Jim Ingalls, Jessica Krick, Bill Glaccum, Sean Carey, William Mahoney, Elena Scire, Elise Furlan, Yi Mei, and Joseph C Hunt, 2019, {\it Spitzer Space Telescope: Innovations and Optimizations in the Extended Mission Era}, 2018 SpaceOps Conference, Space Operations: Inspiring Humankind’s Future, 28 May - 1 June 2018 Marseille, France, (\linkable{https://arc.aiaa.org/doi/10.2514/6.2018-2350})

\subsection{Papers Related to Mission Design and Mission Operations}

\subsubsection{}
Barba, Stephen J, Lisa J Garcia, Douglas B McElroy, David S Mittman, JoAnn C O'Linger, and Steven R Tyler, 2006, {\it Planning and Scheduling the Spitzer Space Telescope}, Observatory Operations: Strategies, Processes, and Systems, International Society for Optics and Photonics, 6270, 62700Z.

\subsubsection{}
Hunt, Joseph C., and Leo Y. Cheng, 2012, {\it Re-Engineering the Mission Operation System (MOS) for the Prime and Extended Mission}, 12th International Conference on Space Operations (AIAA SpaceOps 2012), 1285801.
 
\subsubsection{}
Kwok, Johnny H, Mark D Garcia, Eugene Bonfiglio, and Stacia M Long, 2004, {\it Spitzer Space Telescope mission design}, Optical, Infrared, and Millimeter Space Telescopes, SPIE Astronomical Telescopes + Instrumentation, Glasgow, United Kingdom, International Society for Optics and Photonics, 5487, 201-210, doi:10.1117/12.551598.

\subsubsection{}
Mahoney, William A, Susan Comeau, Lisa J Garcia, Douglas B McElroy, David S Mittman, JoAnn C O'Linger, and Steven R Tyler, 2008, {\it Spitzer Scheduling Challenges: cold and warm, Observatory}, Operations: Strategies, Processes, and Systems II, International Society for Optics and Photonics, 7016, 70161W. 

\subsubsection{}
Mahoney, William A, Lisa J Garcia, Joseph Hunt Jr, Douglas B McElroy, Vince G Mannings, David S Mittman, JoAnn C O'Linger, Marc Sarrel, and Elena Scire, 2010, {\it Spitzer Warm Mission transition and operations}, Observatory Operations: Strategies, Processes, and Systems III, International Society for Optics and Photonics, International Society for Optics and Photonics, 7737, 77371W

\subsubsection{}
Mahoney, William A, Mark J Effertz, Mark E Fisher, Lisa J Garcia, Joseph C Hunt Jr, Vincent Mannings, Douglas B McElroy, and Elena Scire, 2012, {\it Spitzer operations: scheduling the out years}, Observatory Operations: Strategies, Processes, and Systems IV, International Society for Optics and Photonics, International Society for Optics and Photonics, 8448, 84481Z

\subsubsection{}
Sarrel, Marc A, and Joseph C Hunt Jr, 2008, {\it Evaluating requirements on the Spitzer Mission operations system based on flight operations experience}, SPIE Astronomical Telescopes + Instrumentation, Marseille, France, International Society for Optics and Photonics, 7016, 70161P.

\subsubsection{}
Scire, E, 2014, {\it Warm Spitzer: Effects of Major Operational Changes on Publication Rates}, Astronomical Data Analysis Software and Systems XXIII, 485, 481.

\subsubsection{}
Scott, Charles P, Bolinda E Kahr, and Marc A Sarrel, 2006, {\it Spitzer observatory operations: increasing efficiency in mission operations}, SPIE Observatory Operations: Strategies, Processes, and Systems, Bellingham, WA, International Society for Optics and Photonics, 6270, 62701B.

\subsubsection{}
Scott, Charles P, and Robert K Wilson, 2006, {\it Spitzer pre-launch mission operations system: the road to launch}, SPIE Observatory Operations: Strategies, Processes, and Systems, International Society for Optics and Photonics, 6270, 627013.

\subsection{Papers Related to Science Operations}

\subsubsection{}
Dodd, Suzanne R, 2004, {\it The Spitzer science operations system: How well are we really doing?}, Optical, Infrared, and Millimeter Space Telescopes, International Society for Optics and Photonics, 5/13/2004, 5487, 158-165. 

\subsubsection{}
Storrie-Lombardi, Lisa J, 2012, {\it Spitzer Space Telescope: Unprecedented Efficiency and Excellent Science on a Limited Budget}, Astronomical Data Analysis Software and Systems XXI, Astronomical Society of the Pacific Conference Series, No.461, Astronomical Society of the Pacific, San Francisco, CA, 125-134.

\subsubsection{}
Storrie-Lombardi, Lisa J, Bolinda E Kahr, Joseph C Hunt, Sean Carey, Carson Lee Bennett, Nancy Silbermann, Elena Scire, William A Mahoney, and Patrick Lowrance, 2018, {\it Lessons learned in extended-extended Spitzer Space Telescope operations}, SPIE Astronomical Telescopes + Instrumentation, Austin, TX, International Society for Optics and Photonics, 10704, 107041D.

\subsection{Papers Describing In-Orbit Checkout}

\subsubsection{}
Linick, Sue H, John W Miles, CA Boyles, and JB Gilbert, 2004, {\it Spitzer Space Telescope in-orbit checkout and science verification operations}, AIAA SpaceOps 2004 Conference.

\subsubsection{}
Miles, John W, Sue H Linick, Carole Boyles, Mark D Garcia, John B Gilbert, Stacia M Long, Michael W Werner, and Robert K Wilson, 2004, {\it Execution of the Spitzer in-orbit checkout and science verification plan}, Optical, Infrared, and Millimeter Space Telescopes, International Society for Optics and Photonics, 12 October 2004, 5487, 134-145, doi:10.1117/12.552330.

\section* {Acknowledgments}
We thank JATIS editor Mark Clampin for suggesting that we include a discussion of integration and test in the paper. We thank Sean Carey of the {\it Spitzer} Science Center for providing figure 11. We also thank Tom Soifer of the {\it Spitzer} Science Center for his helpful comments on the manuscript, and Wayne Evenson of Lockheed Maritn, Denver, for early versions of figures 3 - 5.  Our colleagues Paul Chodas, Bob Gehrz, Erick Young, and Dave Bayard cheerfully provided answers to numerous questions about {\it Spitzer}. Insightful comments of three referees led to significant improvements in this paper.  The research was carried out at the Jet Propulsion Laboratory, California Institute of Technology, under a contract with the National Aeronautics and Space Administration (80NM0018D0004). This work is based in part on observations made with the {\it Spitzer Space Telescope}, which was operated by the Jet Propulsion Laboratory, California Institute of Technology under a contract with NASA.

The cost information contained in this document is of a budgetary and planning nature and is intended for informational purposes only. It does not constitute a commitment on the part of JPL and/or Caltech.

\section* {Biographies}

\hspace{\parindent} {\bf Dr. Michael Werner} is the Project Scientist of the {\it Spitzer Space Telescope} at the Jet Propulsion Laboratory, California Institute of Technology.  He has also served as the Chief Scientist for Astronomy and Physics at JPL. He has authored and coauthored dozens of papers presenting scientific results from {\it Spitzer} and is the author, with coauthor Peter Eisenhardt, of the book
``More Things in the Heavens: How Infrared Astronomy is Expanding our View of the Universe,'' (Princeton University Press, 2019), which
discusses {\it Spitzer} science in depth.

{\bf Dr. Patrick J Lowrance} is a senior staff scientist at Caltech/IPAC specializing in brown dwarfs and exoplanet science. He is currently part of the ground system team of the Roman Space Telescope Science Support Center. He has worked with different aspects of the science and operations of space-based observatories and instruments including the {\it Hubble} and {\it Spitzer Space Telescopes} for over two decades.

 {\bf Dr. Thomas L. Roellig} has worked in the NASA civil service as an astrophysicist at the NASA Ames Research Center since 1980 and is currently the Chief of the Astrophysics Branch at Ames. His scientific research interests have spanned a wide range of infrared astronomy and astronomical instrumentation development. He has conducted research and published papers in infrared instrument development, solar science, solar system science, star formation, interstellar medium, and brown dwarf astronomy.

{\bf Dr. Varoujan Gorjian} is a Research Scientist at NASA's Jet Propulsion Laboratory, California Institute of Technology. He has been involved with the {\it Spitzer Space Telescope} for over 20 years as both a member of the {\it Spitzer} Project Science office at JPL as well as a scientific user of the telescope. His main astronomical interests are variability in active galactic nuclei, characterization of exoplanets, and detection of the cosmic infrared background.  

{\bf Joseph Hunt, Jr} is a project manager at NASA’s Jet Propulsion Laboratory in Pasadena, Ca.  One of his most passionate past roles was Deputy Mission Manager and Flight Director for NASA’s {\it Spitzer Space Telescope} mission.  He has over forty-two years of diverse aerospace engineering experience with professional expertise in the disciplines of flight simulation, inertial measurement units, Earth orbiting satellites, and interplanetary and deep space spacecraft, flight, ground and mission operations systems and processes.  He currently serves as the project manager for Mars Odyssey and NEOWISE missions.

{\bf Dr. C.M. (Matt) Bradford} obtained his doctorate in Astronomy and instrumentation at Cornell in 2001. He held a Millikan postdoctoral fellowship at Caltech from 2001-2003 and has been on the science staff at JPL since that time. He enjoys developing and fielding new submillimeter and millimeter-wave instrumentation on mountain-top sites, using the datasets to study interstellar medium conditions in galaxies near and far. Current projects include a balloon-borne far-IR spectrometer, an on-chip mm-wave spectrometer demonstration, and a mm-wave line intensity mapping instrument. He is also active in developing concepts and the necessary detector systems for cryogenic space far-IR astrophysics missions.

{\bf Dr. Jessica Krick} is an astronomer and data scientist on the Infrared Science Archive science platform team at Caltech/IPAC.  She received her PhD in Astronomy and Astrophysics from the University of Michigan and has worked as part of the {\it Spitzer} IRAC instrument support team.  Her scientific interests range from zodiacal light to exoplanets to clusters of galaxies and include machine learning and data science techniques applied to big astronomical datasets.

\bibliography{report}   % bibliography data in report.bib
\bibliographystyle{spiejour}   % makes bibtex use spiejour.bst

\copyright{2022} All Rights Reserved

\section* {Figure Captions}

\hspace{\parindent} {\bf Figure 1.} Cutaway view of the {\it Spitzer} observatory\cite{Werne2004}. The dust cover atop the telescope tube was jettisoned a few days after launch, and the cryostat aperture door was opened shortly afterwards to admit infrared radiation into the instrument chamber. See papers listed in the appendix for a detailed discussion of the in-orbit checkout for {\it Spitzer}.

{\bf Figure 2.} Ball Aerospace thermal model\cite{gehrz2007}. Heat input is
solely from insolation on the solar panel. Cooling of the
cryogenic telescope assembly is accomplished by radiation
and vapor cooling. Heat is transferred through the
system along the paths indicated by the arrows by radiation
(dashed blue arrows), conduction (solid green
arrows), and vapor cooling (broad orange arrows). The
equilibrium temperatures for the various observatory
components are given for the case when the cryogenic
telescope is operating at 5.5 K. The model assumes a
focal-plane heat dissipation of 4 mW and an insolation
of 5.3 kW. Courtesy of Ball Aerospace/JPL-Caltech.

{\bf Figure 3.} The assembled {\it Spitzer} observatory being prepared for thermal vacuum testing at Lockheed Martin in Sunnyvale, CA.  The photograph illustrates the key features of the thermal control system up to and including the outer shell of the telescope. These include the chevron-shaped solar panel, the solar panel shield immediately behind it, and the telescope outer shell, with its anti-solar side painted black to maximize its infrared emissivity.  Also visible between the spacecraft and the outershell is the spacecraft shield which isolated the outer shell from the spacecraft. Credit: NASA.

{\bf Figure 4.} The average temperature of the lower, solar cell-containing portion of the solar panel vs. time over the entire {\it Spitzer} mission.  The annual periodicity seen in this and the other temperature data reflects the eccentricity of {\it Spitzer’s} orbit.

{\bf Figure 5.} Spitzer outer shell temperature vs. time over the entire {\it Spitzer} mission.  The discontinuity in 2009 reflects the exhaustion of the liquid helium and the start of the Warm Mission, when vapor cooling was no longer available. The oscillations in temperature seen before this reflects the variation in vapor cooling resulting from the helium utilization strategy described in the text.

{\bf Figure 6.} Spitzer primary mirror temperature vs. time over the entire {\it Spitzer} mission.  The fluctuations in temperature seen prior to the start of the Warm Mission in mid-2009 reflect the helium utilization strategy described in the text.

{\bf Figure 7.} Achieved {\it Spitzer} point source sensitivities (solid lines, 1$\sigma$ in 500 sec)\cite{Werne2004} compared with estimates of fundamental limits imposed by the zodiacal background. The short vertical lines under each instrumental bandpass show the zodiacal  background-limited NEFD estimate for the bandpass, assuming noise from the zodiacal light in an annual average sightline viewing the north ecliptic pole. Note that the achieved performance is closest to the background limit in the 8-to-24$\mu$m range were the background is at is brightest. The background couples in both polarizations with a square bandpass of the plotted width and total transmission (including detector absorption efficiency) of 0.44, 0.42, 0.14, and 0.30 for IRAC bands 3.6, 4.5, 5.8, and 8.0 $\mu$m,  0.6, 0.18, and 0.15 for MIPS 24, 70 and 160 $\mu$m, and 0.65 for the IRS blue peak-up array (PU B) at 16 $\mu$m. Perfect detectors which add no noise other than generation-recombination noise are assumed throughout. The sensitivity achieved by MIPS is degraded below the zodiacal limit, in part, by the effects of confusion at both 70 and 160 $\mu$m; it is appreciable in both channels in 500 sec.   The IRS short-low and long-low module sensitivities are referred to a $\lambda / \delta \lambda$ = 50 bin; the estimates adopt wavelength-independent efficiencies of 12\% and 8.5\%, respectively; this is meant to include all sources of loss including slit coupling, blaze efficiency in both polarizations, filters and detector quantum efficiency.   For all, it is assumed that the 14\% obscured 85-cm {\it Spitzer} telescope (A$_{\rm geom}$ = 0.489~m$^2$) couples to a point source with 75\% aperture efficiency.   The $\sqrt{2}$ photo-electron recombination penalty is included for all bands except IRAC 1 and 2, where it does not apply.

{\bf Figure 8.} Average number of pixels in the IRAC arrays affected by cosmic rays in the calibration frames taken during the {\it Spitzer} Warm Mission\cite{lowrance2018}.  The array formats were 256$\times$256 30$\mu$m-squared pixels,  The spike in 2012 was due to a solar flare.

{\bf Figure 9.} Average IRAC responsivity vs. time for the final 8 years of the Warm Mission, based on the measured flux from seven primary calibrator stars. The seven primary calibrators for {\it Spitzer} were KF09T1,KF06T2, KF08T3, KF06T1, NPM1p67, NPM1p60,
1812095\cite{reach2005}.

{\bf Figure 10.} Average solar panel-generated current over time for the entire {\it Spitzer} mission.

{\bf Figure 11.} Comparison of pointing repeatability using peak-up vs. blind pointing.  Each point represents the mean centroid of the first 64 frames of the following:  78 observations (dark circles) beginning with PCRS Peak-Up followed by an offset to the 4.5 $\mu$m sweet-spot using gyros; 20 observations (light squares)  using PCRS Peak-Up and the star tracker for the offset; and 48 observations (light triangles) initiated  without PCRS peakup. Each set of symbols has had the 
mean position of the set subtracted.  An 0.5$\times$0.5 pixel square is overlaid on the plot, indicating the size of the sweet-spot region\cite{Ingalls2012}.

{\bf Figure 12.} Effects of tightening battery heater dead band on pointing and photometric stability.  The upper two panels show the variation of the (x,y) position of the centroided stellar image on the array, while the lower panel shows the variation in measured signal, all as a function of time.  Black points are data taken before the dead band was tightened up; green points were taken after the adjustment.

\end{spacing}
\end{document}